%% file: main.tex
\documentclass[manuscript]{acmart}

\AtBeginDocument{%
  \providecommand\BibTeX{{%
    \normalfont B\kern-0.5em{\scshape i\kern-0.25em b}\kern-0.8em\TeX}}}

\setcopyright{acmcopyright}
\copyrightyear{2022}
\acmYear{2022}
\acmJournal{PACMHCI} %tiis
\acmVolume{00}
\acmNumber{00}
\acmArticle{00}
\acmMonth{00}

\acmSubmissionID{9012}

\usepackage{pdflscape}
\usepackage{subcaption}
\usepackage{rotating}
\usepackage{graphicx}
\usepackage{comment}
\usepackage{todonotes}
\usepackage{soul}
\usepackage{afterpage}
\usepackage{array}
\newcolumntype{C}{>{\global\let\currentrowstyle\relax}}
\newcolumntype{^}{>{\currentrowstyle}}
\newcommand{\rowstyle}[1]{\gdef\currentrowstyle{#1}%
  #1\ignorespaces
}

\newcommand{\rv}[1]{{\leavevmode\color{black}#1}} %revision
\newcommand{\mr}[1]{{\leavevmode\color{blue}#1}} %camera-ready

\renewcommand{\mr}[1]{{\leavevmode\color{black}#1}} %camera-ready
\renewcommand{\st}[1]{}
\begin{document}

%\title{Investigating the Effects of Timing and explanations of AI-Assistance on UX Evaluators' Perception and Performance}

\title{Human-AI Collaboration for UX Evaluation: Effects of Explanations and Synchronization}
%\title{Human-AI Collaboration for UX Evaluation: Effects of Work Context and Agency}

\author{Mingming Fan}
\authornote{Corresponding author}
\email{mingmingfan@ust.hk}
\affiliation{%
  \institution{The Hong Kong University of Science and Technology}
   %\city{Hong Kong SAR and Guangzhou}
\country{China}
}

\author{Xianyou Yang}
\authornotemark[2]
\email{xy1258@rit.edu}
\affiliation{%
  \institution{Rochester Institute of Technology}
\country{USA}
}

\author{Tsz Tung Yu}
\email{tt3yu@uwaterloo.ca}
\affiliation{%
  \institution{University of Waterloo}
\country{Canada}
}

\authornote{equal contribution}

\author{Q. Vera Liao}
\email{veraliao@microsoft.com}
\affiliation{%
  \institution{Microsoft Research}
  \city{Montreal}
\country{Canada}
}

\author{Jian Zhao}
\email{jianzhao@uwaterloo.ca}
\affiliation{%
  \institution{University of Waterloo}
\country{Canada}
}

% \affiliation{%

% \affiliation{%

% \affiliation{%

% \affiliation{%

% \affiliation{%



\renewcommand{\shortauthors}{Fan, et al.}

\begin{abstract}
Analyzing usability test videos is arduous. Although recent research showed the promise of AI in assisting with such tasks, it remains largely unknown how AI should be designed to facilitate effective collaboration between user experience (UX) evaluators and AI.
Inspired by the concepts of \textit{agency} and \textit{work context} in human and AI collaboration literature, we studied two corresponding design factors for AI-assisted UX evaluation: \textit{explanations} and \textit{synchronization}. 
Explanations allow AI to further inform humans how it identifies UX problems from a usability test session; synchronization refers to the two ways humans and AI collaborate: \textit{synchronously} and \textit{asynchronously}.
We iteratively designed a tool---AI Assistant---with four versions of UIs corresponding to the two levels of explanations (with/without) and synchronization (sync/async). By adopting a hybrid wizard-of-oz approach to simulating an AI with reasonable performance, we conducted a mixed-method study with 24 UX evaluators identifying UX problems from usability test videos using AI Assistant.
Our quantitative and qualitative results show that AI with explanations, regardless of being presented synchronously or asynchronously, provided better support for UX evaluators' analysis and was perceived more positively; when without explanations, synchronous AI better improved UX evaluators' performance and engagement compared to the \mr{asynchronous} AI.  
Lastly, we present the design implications for AI-assisted UX evaluation and facilitating more effective human-AI collaboration.
\end{abstract}




\maketitle

\input{texfiles/1-introduction}
\input{texfiles/2-relatedwork}

\input{texfiles/3-method}

\input{texfiles/4-userstudy}
\input{texfiles/5-quantitativeresults}
\input{texfiles/6-qualitativeresults}

\input{texfiles/7-discussion}
\input{texfiles/8-conclusion}

% \begin{acks}

% \end{acks}

\bibliographystyle{ACM-Reference-Format}
\bibliography{main}

% \appendix

\end{document}

%% file: texfiles/1-introduction.tex
\section{Introduction}

Usability testing has been widely adopted in both industry and academia to identify user experience (UX) problems when developing interfaces and systems~\cite{mcdonald2012exploring,fan2020Survey}.  
With the convenience and easy access to diverse users, remote usability testing, by conducting and recording the tests through video conferencing tools and analyzing the recorded sessions afterwards,  has become increasingly popular~\cite{andreasen2007happened,fan2020Survey,castillo1998remote}.
%which makes it easier than ever to carry out and record a large number of usability test sessions.
%Although this has made it possible to conduct large amounts of remote usability test sessions, 
However, analyzing recorded usability test sessions is labor-intensive and time-consuming, as UX evaluators have to listen to what users verbalize while watching their actions to catch sporadic problems in an often lengthy video. %and sometimes the videos can be long. %it typically requires reviewing long session videos.
%On the other hand, recent research has shown a great promise of Artificial Intelligence (AI) technologies in facilitating user-centered research processes, such as qualitative coding ~\cite{}.
In practice, UX evaluators often have to complete the analysis in a short period~\cite{folstad2012analysis} \rv{and tend to perform quick rather than thorough analysis}. 
\rv{What's more, UX evaluators who analyze the same usability test sessions have been found to identify substantially different sets of usability problems, which is known as the ``evaluator effect''~\cite{hertzum2001evaluator}}.
\rv{One important way to cope with
the ``evaluator effect'' is to involve multiple
evaluators to analyze the same test session.}
\rv{However, due to practical constraints (e.g., limited company resources), few evaluators (23\%) had a chance to gain a different perspective from other \textit{human} evaluators on the same usability test session~\cite{folstad2012analysis}.}
\rv{One possible solution, in light of recent advancement in AI, is to investigate whether an AI agent could be designed to analyze the same usability test session and provide a different perspective to the human  evaluator}.

%\vera{to-do: @Mingming The second paragraph needs to be re-written: 1) clarify on the challenge of AI assisting usability video analysis (per 2AC's comments it is not just a "video annotation" task), 
%2) the design is motivated by the needs of the video analysis task rather than just CSCW theories (why explanation? why synchronization? can you cite prior work?)--then discuss this dimensions correspond to dimensions in human-human collabroation}

% \ming{@Vera: DONE. Please Check and update my revisions and comment out these comments.}

%As recent research has shown that Artificial Intelligence (AI) technologies could improve the efficiency of user-centered research, such as coding~\cite{}, it is worth explore how to design AI-assisted tools to assist UX evaluators with analyzing usability test videos.
%especially AI for text mining, computer vision, and audio analysis, has been exploited to assist with UX evaluation, 
%Although artificial intelligence (AI) has been leveraged to detect UX problems \cite{grigera2017automatic,fan2020detection,paterno2017customizable,harms2019automated,jeong2020detecting}, fully automatic detection of UX problems is still in its infancy. To address the limitation,
\rv{Toward this goal, recently} researchers began to explore human-AI collaborative approaches, in which AI \rv{plays the role of another evaluator and} suggests potential \rv{usability} problems to the UX evaluator. 
\rv{%One representative work is VisTA, an analytical tool that visualizes an AI's predicted problems to a UX evaluator before and during their analysis~\cite{fan2020vista}. In VisTA, the AI finishes its analysis before the evaluator starts to analyze the same test session. 
For example, VisTA~\cite{fan2020vista} is an analytical tool that visualizes AI predicted problems before a UX evaluator starts their analysis; thus, the collaboration between the AI and the evaluator is \textit{asynchronous}. Such asynchronous human-AI collaboration was also explored in other contexts~\cite{CoUX2021TVCG,sarvghad2015exploiting,sarvghad2017visualizing,xu2018chart,zhao2017supporting,willett2011commentspace,heer2007voyagers,wattenberg2006designing}. In human-human collaboration, synchronization, whether the collaboration happens \textit{asynchronously} and \textit{synchronously}, is considered a key dimension that impacts the user requirements and design of CSCW systems, as seen in the \textit{Time} dimension of Johansen's Time-Space Matrix~\cite{johansen1988groupware}. To our knowledge, \textit{synchronous} collaboration between humans and AI is relatively understudied in knowledge-rich domains; What's more, no research has compared asynchronous and synchronous collaboration in the context of usability evaluation. We took a first step to investigate how \textbf{synchronization} between the AI and UX evaluators might affect their collaboration experiences.}
%We believe the understanding of this problem could inspire the design of human-AI collaboration in other knowledge-rich domains in addition to usability evaluation.} 

\rv{Furthermore, when analyzing a usability test video, UX evaluators care about not only \textit{where} in the video the AI believes the user encounters a problem but also \textit{why} it believes so--the \textbf{explanations} of AI's judgment. Such an explanation could allow evaluators to better assess whether to incorporate or reject AI's judgement~\cite{fan2020vista}. 
%For example, in the user study of VisTA~\cite{fan2020vista}, participants expressed a desire to see not only \textit{what} input features contributed to AI's judgment, but also \textit{why} these features are indicators of usability problems. 
On the other hand, recent work on ``agency divide'' between human and AI~\cite{lai2019human,Lai2020Why} suggests that providing explanations could affect human's perceived agency. Specifically, humans perceive full agency of their work when there is no AI assistance, and decreasing agency as the amount of information and assistance provided by the AI increases, including explanations. A decreased human agency could negatively affect the engagement and outcome of usability video analysis. Given these potential effects, we empirically investigated how \mr{accessing} AI's explanation, and how explanation and synchronization working together, affect human-AI collaboration in the context of usability test video analysis. }

 Towards these study goals, we iteratively designed \textit{AI Assistant}, a tool assisting UX evaluators with analyzing usability test videos. 
 As it is still challenging to develop an AI that could detect UX problems of any given usability test video with consistent accuracy, we adopted a wizard-of-oz (WoZ) approach to simulating an AI that could detect UX problems with reasonable precision and recall. \rv{We also designed WoZ explanations based on both what matters in analyzing usability test video~\cite{fan2020vista} and what is potentially feasible with state-of-the-art explainable AI (XAI) techniques.}
 We designed AI Assistant with four different versions of user interfaces (UIs), which present the AI-suggested \rv{usability} problems \textit{synchronously} or \textit{asynchronously} and \textit{with} or \textit{without} explanations. 
 We conducted a mixed-methods study with 24 UX evaluators, in which they analyzed two usability test videos with AI Assistant and were interviewed afterwards regarding their experiences, perceptions, and preferences of AI Assistant.
 %The controlled user study adopted a mixed factorial design, where synchronization was the within-subject factor, and explanations was the between-subjects factor. 
 
 Our results showed that both synchronization and explanations positively affected human-AI collaboration.
 Compared to the asynchronous AI without explanations, AI with explanations, regardless of being presented asynchronously or synchronously, helped to improve UX evaluators' performance of (e.g., the number of identified UX problems) and engagement (e.g., time spent) in their analysis, and their perception of AI Assistant (e.g., understanding of AI).
 When without explanations, the synchronous AI helped to improve UX evaluators' performance and engagement more than the asynchronous AI.
Our qualitative results from semi-structured interviews provide further insights into the effects of the two factors. Based on the findings, we discuss the design implications of explanations and synchronization for UX evaluation and effective human-AI collaboration.

%% file: texfiles/2-relatedwork.tex
\section{Background and Related Work}

\subsection{AI for \rv{Usability} Problems Detection}

Analyzing usability test video to identify problems that participants encountered is a common task for UX evaluators. \st{Such analysis differs from a video annotation task commonly adopted in computer vision or crowdsourcing research, which typically requires workers to select a label from a predefined list. There is no predefined set of problems for UX evaluators to select from. Instead,} \mr{They need to attend to multiple behavioral signals of the participant from both visual and audio channels of the test video. More importantly, they need to leverage their domain expertise to determine whether there is indeed a usability problem or whether it is just typical efforts (e.g., trial-and-error) that the participant had to make when using the product.}  
\rv{To help UX evaluators better analyze usability test videos, }
recent research began to develop AI methods to predict the overall UX of interfaces~\cite{oztekin2013machine} or detect specific \rv{usability} problems~\cite{grigera2017automatic,paterno2017customizable,harms2019automated,jeong2020detecting}.  
For example, Grigera et al. proposed a rule-based classifier to detect a predefined set of usability smells---the hints of bad designs that could cause usability problems---by analyzing users' interaction logs~\cite{grigera2017automatic}. 
Similar rule-based classification approaches have been used to detect usability smells in mobile websites~\cite{paterno2017customizable} and in virtual reality applications~\cite{harms2019automated}.
Jeong et al. proposed a graph-based AI method to model and measure the similarity of users' interactions with a mobile application to detect potential usability problems~\cite{jeong2020detecting}.
Although such automatic methods show the promise of detecting simple \rv{usability} problems for specific user interfaces based on structured data (e.g., interaction logs) and empirical rules, detecting \rv{usability} problems directly from a usability test video, which is a primary task of UX practitioners, is still challenging to be fully automated as it requires \mr{an} understanding of the functions and designs of the test product, the test tasks, and the test subject's behaviors.

%\vera{to-do: @Mingming, can you revise this paragraph to call out limitation of VisTA (R2) and how it motivates explanation and synchronization? you can decide what goes here v.s. intro}

%\ming{@Vera: DONE. Please Check and update my revisions and comment out these comments.}

To address the limitation of fully automated methods, researchers began to investigate human-AI collaboration tools to support UX evaluators rather than replacing them. 
%automatic detection  these methods were not designed to support the universal workflow and jobs of UX professionals, in which a primary task is to analyze usability test videos to identify UX problems. 
%These days it is common for UX professionals to perform such tasks remotely, then analyze the recorded videos to identify usability problems. 
One representative work is VisTA, a visual analytical tool that integrates visualization and machine learning (ML) to detect and highlight segments of a usability test video containing potential \rv{usability} problems~\cite{fan2020vista}. \rv{Such human-AI collaboration was shown to help} UX evaluators identify more problems \rv{than they worked alone~\cite{fan2020vista}.}  \rv{As a first step to experiment human-AI collaboration for usability test video analysis, VisTA was limited in two ways. First, the collaboration between the AI and the UX evaluator was \textit{asynchronous}. The evaluator was shown with the AI's predictions before they started their analysis, which might have affected the evaluator's independent analysis. Alternatively, the AI's predictions could be shown \textit{synchronously}, creating a perception that the evaluator and the AI are analyzing a usability test video simultaneously, which might allow the evaluator to perform more independent analysis while still being able to access the second perspective from the AI.} 

\rv{The second limitation, based on their study participants' feedback, was that VisTA remained an ``opaque box'' and did not allow an understanding of how the AI worked. Although VisTA visualizes the input features to show \textit{what} went into the AI's analysis, it was deemed insufficient without \mr{having} access to the meaning of these input features or \mr{the rationales on which the judgments were based}. Thus, it is necessary to explore explanations that could answer the \textit{why} \mr{question} in identifying usability problems in a video, such as what design principles or usability heuristics were violated~\cite{fan2020vista}.}

%\st{from usability test sessions with the support of VisTA}. \st{Although VisTA demonstrated the promise of human-AI collaboration for UX evaluation, it adopted an asynchronous collaboration between UX evaluators and AI. It remains unknown how to design human-AI collaboration to best support the analysis of usability test videos. }
%However, VisTA relies on supervised ML that needs to be trained on a set of usability test videos on the same or similar products with labels of usability problems~\cite{fan2020vista}. Thus, it remains a challenge to develop AI or ML methods that could reliably detect UX problems in usability tests across different systems. 

%\st{Our work intends to study AI assistants that not only offer similar functions as VisTA--suggesting segments with potential UX problems in a recorded usability test video---but also collaborate with UX evaluators in different ways.} 
\rv{Our work seeks to understand how synchronization (i.e., synchronous and asynchronous collaboration) and explanations (i.e., with and without) would affect UX evaluators' collaboration with the AI in the context of usability test video analysis.} To overcome the technical limitations of AI in detecting UX problems, we opt for a hybrid Wizard-of-Oz (WoZ) approach in order to focus on evaluating different ways to present the AI's suggestions with a controlled experiment. Our approach simulates a reliably performing AI system but is grounded in the technical feasibility of ML technologies. Specifically, we assume the system works with a supervised ML classifier, which takes a segment of a usability test video as input and predicts whether the test subject encounters a problem in that video segment. The design of our WOZ AI's functionalities, including the AI explanations, will be discussed in Section~\ref{sec:wozAI}.

\subsection{Human-AI Collaboration}
The term ``human-AI collaboration'' has emerged in recent research studying the usage of AI systems~\cite{arous2020opencrowd,wang2019human,cai2019hello}. 
There is a shift from an ``automation'' perspective of AI to recognizing that AI systems should be used to support, instead of replacing, the decisions and tasks of domain workers. 
For example, an ``algorithm-in-the-loop'' process~\cite{green2019principles} is adopted in many high-stakes decision-making contexts, where ML models are used to inform people its assessment or to alert cases falling in a targeted category. 
Ultimately, people have to decide whether to accept the AI's recommendations or discount them~\cite{zhang2020effect}. 
Following this collaborative perspective, researchers studied how people perceive algorithmic decision assistance~\cite{binns2018s,lee2018understanding}, how they use AI systems for decision-making~\cite{cai2019hello,cai2019human}, often in unexpected or suboptimal ways~\cite{yang2019unremarkable,christin2017algorithms}, and how to improve the human-AI joint decision-making outcomes~\cite{bansal2019updates,wilder2020learning,zhang2020effect}.

While research has only begun to define and study different types of human-AI partnership, it would be advisable to draw on theories and insights from human-human collaboration research. CSCW research has identified a number of core dimensions to characterize different collaborative tasks. 
One is the division of work and labor between parties~\cite{strauss1985work}. 
%In parallel, 
Recent work by Lai et al.~\cite{lai2019human,Lai2020Why} proposed a notion of ``divide of agency'' between human and AI with a spectrum from full human agency to full automation. 
%Focusing on the utility of AI explanations, given recent development in explainable AI (XAI) that enables AI to provide rationales for its decisions, 
% While the AI that provides both predictions and explanations is considered to reduce human \textit{agency}, the AI that provides only predictions without any explanations is considered to have lower \textit{machine agency}. 
%that AI providing predictions alone. 
%The increase of machine agency does not only hinge on the perception of competence and trustworthiness engendered by explanations~\cite{pu2006trust,ribeiro2016should}, but also in that the explanations itself could provide additional assistance for decision-making, for example by highlighting important factors in the decision or illuminating useful rationales~\cite{lai2019human,Liao2020Question}. 
In particular, AI that provides \textit{explanations} for its predictions is considered to \rv{allow lower human agency} than one that offers predictions only. On the other hand, explanations not only could affect the perception of AI's competence~\cite{pu2006trust,ribeiro2016should}, but also provide additional assistance for decision-making, for example, by highlighting important factors in the decision or illuminating useful rationales~\cite{lai2019human,Liao2020Question}. 
%In this research, we conduct an experiment that compares the AI assistant that provides predictions \textit{with} and \textit{without} \textit{explanations} that would change human agency to understand the effect of explanations on UX evaluators' performance and preference when analyzing usability test videos.
\mr{However, such improvements from explanations were typically observed with the AI that outperformed the human. Bansal et al. recently examined how the explanations of the AI with human-comparable performance affect human-AI collaboration \cite{bansal2021does}. Our work extends theirs in two aspects. First, in Bansal et al.'s studies, AI’s explanations were presented to human participants upfront before they started their own analysis. Such asynchronous collaboration, as they acknowledged, made it almost impossible for the participants to reason independently. In contrast, our work studies the potential effects of synchronization between humans and the AI.  
Second, Bansal’s studies used tasks amenable to crowdsourcing, and the findings might not be generalizable to experts in high-stake scenarios. In contrast, our tasks require domain expertise in user experience (UX) research. Thus, the type of human-AI collaboration in our work is between AI and domain experts, and the findings are complementary to those of Bansal et al.’s}.

 The WoZ explanations of AI assistant in this research is informed by the methods developed in the field of Explainable AI (XAI); A full review of prior work on this topic is beyond the scope of this paper but can be found in many recent survey papers~\cite{guidotti2018survey,adadi2018peeking,carvalho2019machine,gilpin2018explaining,arya2019one}. 
Explanations could be generally categorized into ~\textit{global} and \textit{local} explanations. While global explanations describe how a model makes a decision in general, such as how it weighs different features and what rules it follows, local explanations focus on justifying a decision made for a particular instance, for example, by highlighting important features of the instance that contributed to the AI's decision~\cite{arya2019one,guidotti2018survey}. 
The explanations presented in our AI assistant is an example of local explanations, which highlights exceptional features in the usability test video that AI assistant considers as the indicators of a \rv{usability} problem. \rv{To make AI explanations more accessible to \mr{laypeople}, an emerging area of XAI work explored \textit{explanation generation} using domain-specific semantics~\cite{kim2018interpretability,hind2019ted} or rationales~\cite{ehsan2019automated},  enabled by additional human supervision or training data. For example, trained, on human explanation data, rationale generation~\cite{ehsan2019automated} translates AI's internal representation into natural-language rationales.}

Another pivotal dimension to characterize collaborative tasks is the \rv{\textit{Time} dimension} of the well-known Time-Space matrix of human-human collaboration~\cite{johansen1988groupware}. The Matrix characterizes collaborative tasks and computing tools to support workers by a \textit{Time} dimension---whether individuals collaborate \textit{synchronously} or \textit{asynchronously}, and a \textit{Space} dimension---whether collaboration is co-located or geographically distributed.
\rv{While the Space dimension is irrelevant to human-AI collaboration, the Time dimension is pertinent.} However, to our knowledge, little is understood about how human-AI collaboration is impacted by the Time dimension. 
However, these factors could produce distinct interaction experiences that may impact how one perceives and interacts with AI, and ultimately the collaborative outcomes. 
%In this research, besides explanations, we also explore the time axis of work context---\textit{synchronization}---to understand UX evaluators' experiences of working with a \textit{synchronous} and \textit{asynchronous} AI assistant. 
While our current research is carried out in the context of UX evaluation, \textit{synchronization} is broadly applicable to AI-assisted decision tasks.

\subsection{Collaborative Analysis Tools}
Previous research has investigated ways to support collaboration between humans, including design principles (e.g., time and space model~\cite{johansen1988groupware,isenberg2011collaborative} and information scent~\cite{pirolli1999information}), collaborative data analysis and sensemaking \cite{cook2005illuminating,isenberg2011collaborative,Paul10} as well as collaborative infrastructures and tools. 
Examples of collaborative infrastructures include ManyEyes \cite{viegas2007manyeyes} for sharing and commenting on data charts, Polychrome \cite{badam2014polychrome} for collaborative web visualizations, and BEMViewer \cite{mcgrath2012branch} for supporting branch-explore-merge protocol in data exploration.
Collaborative tools support human-human collaborations for synchronous and asynchronous tasks.  Synchronous collaborative tools focus primarily on increasing coworkers' awareness \cite{zhao2017annotation}, building common ground \cite{mahyar2014supporting}, and sharing/constraining personal workspaces \cite{isenberg2010exploratory,tobiasz2009lark}.
Asynchronous collaborative tools often address the design challenges in facilitating communication \cite{heer2007voyagers,wattenberg2006designing}, supporting handoff \cite{zhao2017supporting}, reasoning actions \cite{willett2011commentspace}, and viewing analysis histories \cite{sarvghad2015exploiting,sarvghad2017visualizing,xu2018chart}.

Previous research has primarily focused on supporting collaboration between humans in non-UX domains. In this research, we draw inspirations from this body of literature to design an AI-assisted tool to support synchronous and asynchronous collaboration between AI and UX evaluators when they analyze usability test videos.
This tool is then used as the vehicle to study the effect of the two factors---synchronization and explanations---on UX evaluators' workflow.

%% file: texfiles/3-method.tex
\section{Method}
The goal of this research is to understand the effect of synchronization and explanations on human-AI collaboration for UX research in the context of analyzing usability test videos.
The findings would provide design implications for the two factors in human-AI collaboration.

\subsection{Research Questions}
%To understand the effect of the synchronization and explanations for presenting the AI on the perception and performance of UX evaluators, 
To achieve the above overarching goal, we seek to answer three research questions (RQs):

\begin{itemize}
\item[\textbf{RQ1:}] How would the \textit{explanations} of AI affect UX evaluators' performance and perception in analyzing usability test videos?
\item[\textbf{RQ2:}] How would the \textit{synchronization} of AI affect UX evaluators' performance and perception in analyzing usability test videos?
\item[\textbf{RQ3:}] What are UX evaluators' preferences %\st{for designing \textit{synchronization} and \textit{explanations} of} 
\rv{that can inform the design of} AI tools assisting UX evaluation?
%\item {RQ3}: How would the description and the timing of AI-assistance affect the domain workers' workflow (i.e., analysis process)?
\end{itemize}

\afterpage{
\begin{landscape}
    \begin{figure*}
    % \begin{sidewaysfigure}
      \centering
      \includegraphics[width=\linewidth]{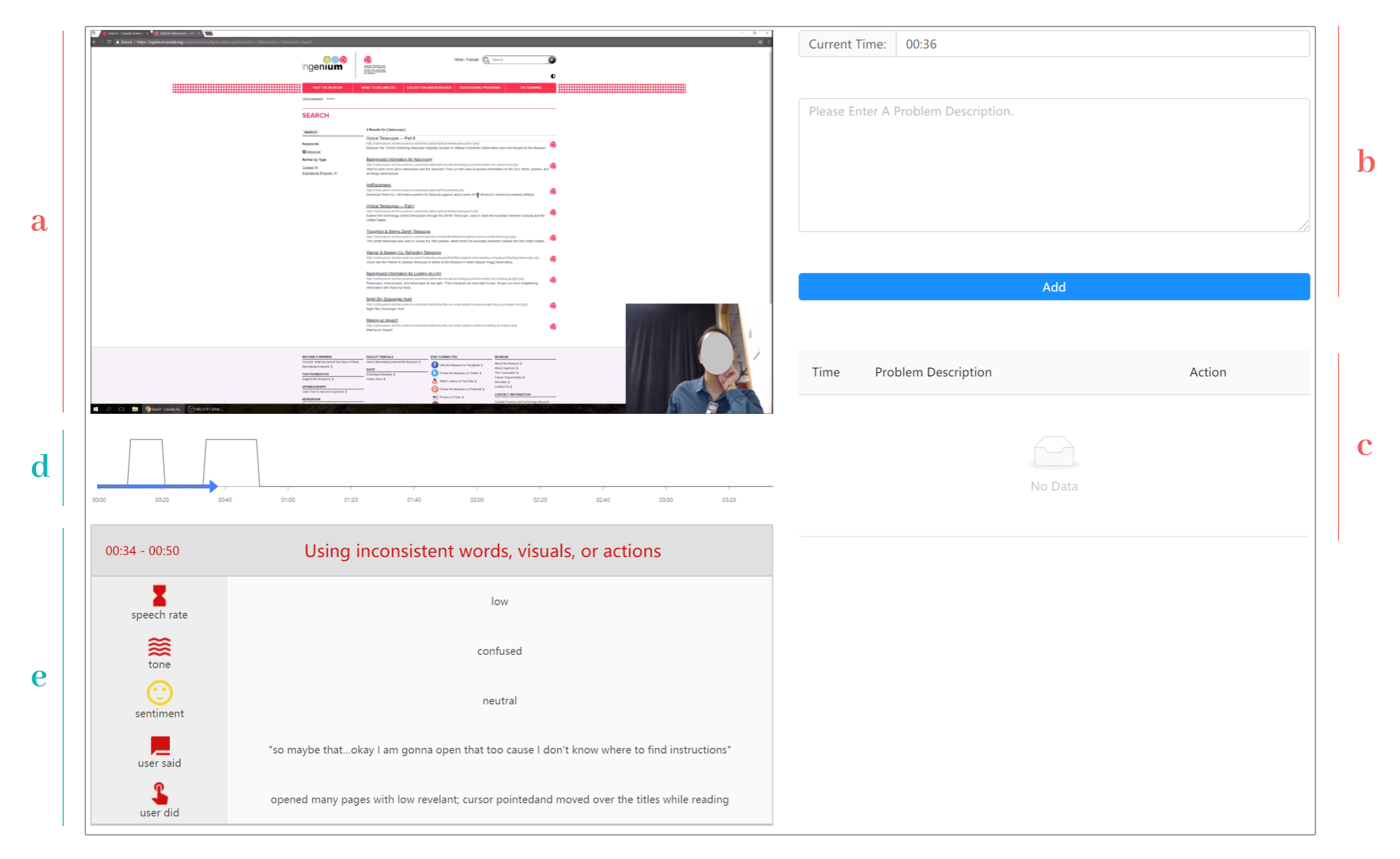}
     \caption{The user interface of AI Assistant that presents the AI-suggested problems \textit{synchronously with explanations}: (a) video player; (b) annotation panel; (c) identified problem table; (d) timeline of the AI-suggested problems; (e) explanations of the AI-suggested problems.}
      \label{fig:UI-Sync-W}
    %  \end{sidewaysfigure}
    \end{figure*}
\end{landscape}
}

\subsection{\textit{AI Assistant} for Usability Test Video Analysis}
Usability test video analysis requires UX evaluators to review the video recordings of test sessions to identify UX problems. We designed a tool---\textit{AI Assistant}---to support the task by suggesting potential UX problems in the video. 
% \sout{Figure~\ref{fig:UI-Sync-W} shows the user interface (UI) for one of the four versions of the AI Assistant (see the supplementary for the rest three versions). The different versions of the AI Assistant present the AI based on the two factors of interest---\textit{explanations} and \textit{synchronization}, which will be discussed in detail in Section~\ref{design-sync-exp}.} 
It supports watching a usability test video, identifying and annotating UX problems, and viewing AI assistance in one place, with the following UI components:

%in the \textit{sync} and \textit{with} explanations condition. 

%Across the four UIs of the AI Assistant, there are four \textit{common} UI components: a video player, an annotation panel, a problem table, and a timeline showing the AI's predictions of the UX problems.

\textbf{Video Player.} AI Assistant includes a basic video player, with which UX evaluators can play, pause, rewind, and jump forward a video (Figure~\ref{fig:UI-Sync-W}a).

\textbf{Annotation Panel.} After identifying a usability problem from watching the test video, UX evaluators often need to describe the problem for reporting purpose or future reference. AI Assistant includes an annotation function, which shows the current timestamp of the video and contains a text field for writing problem description.
Since the annotation function would be used frequently, we position it side-by-side with the video (Figure~\ref{fig:UI-Sync-W}b).
 
\textbf{Problem Table.} To allow UX evaluators to view the problems that they have identified, AI Assistant shows all the problems and the corresponding descriptions in a table (Figure~\ref{fig:UI-Sync-W}c).   

\textbf{AI-Suggested Problems Timeline.} One key function of AI Assistant is introducing AI to assist UX evaluators with their analysis. 
Specifically, the AI analyzes the video and infers whether and when (i.e., a time duration) the user in the video is encountering a usability problem.
%The inferences are represented by two values: \textit{one} means that the user is encounters a problem and zero means that the user does not encounter a problem.
The AI-suggested UX problems are visualized as a horizontal timeline chart, with the value \texttt{1} representing a problem being encountered and the value \texttt{0} representing no problem being encountered (Figure~\ref{fig:UI-Sync-W}d).
%The horizontal line represents the time in the video, and the vertical line represents the AI's inferences.
The timeline is positioned below the video to facilitate UX evaluators to glance at it during their analysis. Detailed design of the timeline will be introduced in Section~\ref{design-sync-exp}.

\textbf{AI Explanations Panel.} For the versions of AI Assistant with explanations, the tool features a separate panel to show the explanations of why the AI infers there is a UX problem in a given time duration.
The explanations is shown below the timeline (Figure~\ref{fig:UI-Sync-W}e). The content and design of the explanations will be described in Sections~\ref{usability-test-videos} and ~\ref{design-sync-exp}.

%In this research, we aim to understand the effect of two aspects of the AI on UX evaluators' perception and performance: Synchronization and explanations.
%Next, we present the unique designs for the four UIs of the AI Assistant regarding the Synchronization and explanations factors (Figure~\ref{fig:UI-Sync-W}e).

%\subsubsection{Hybrid Wizard-of-Oz AI}
As discussed, given the technical challenge to train an ML model that maintains consistently satisfying performance for different usability test videos, we utilized a \textit{wizard-of-oz (WoZ)} approach to generating AI-suggested problems, serving as the back-end AI of the tool.
%Although recent research has shown the possibility of automatically detecting UX problems with some level of accuracy~\cite{fan2020vista,johanssen2019toward,ribeiro2019usability}, it is still challenging to design an AI that could maintain consistently satisfying accuracy for different usability test videos. 
An WoZ approach also allowed us to control the content of explanations to make them consumable for UX evaluators.

%as it remains unknown whether current AI technologies are capable of generating

%human-consumable explanations for UX problems.
%it is unknown whether current UX problem detection algorithms could generate a human-consumable explanations. 
%Therefore, we decided to use a hybrid wizard-of-oz (WoZ) approach to generating AI suggested UX problems and explanations. 
%Consequently, to better focus on the RQs, we decided to adopt the WoZ approach, which allows for better control over the AI-suggested problems and the corresponding explanations.
%Specifically, we inform the problem detection performance and explanations by current AI technologies, algorithmically extracting the model features for the explanations, but manually controlling the suggested problems and create the explanations to better focus on the RQs. 

\subsection{Task Videos and WoZ AI} \label{usability-test-videos}

\subsubsection{Task Videos}
We needed usability test videos for UX evaluators to analyze in our experimental study. 
To do so, we selected two recorded videos with audio from a dataset curated in our previous studies~\cite{fan2019concurrent}.
To create the dataset, we recruited eight native-English speakers (four females and four males, aged 19-26) to participate in usability test sessions, in which they performed tasks on two websites and two physical devices while thinking aloud.
All test sessions were video and audio recorded.
%The two selected videos were about two users using a website and a physical device respectively. 
One of the selected videos was about a participant browsing a website of a national science museum. 
The task was to find a photo of the instructions to operate an early telescope. 
The other selected video was about a participant using a multi-function coffee machine, and the task was to program the machine to make two cups of strong-flavored drip coffee at seven thirty in the morning. 
The website video lasted 214 seconds, and the coffee machine video lasted 682 seconds.
We intentionally selected videos with different interfaces (virtual and physical) as well as different lengths (short and long) to cover a range of scenarios.
Each video was seen by participants as one evaluation task, presented in the video player panel in Figure~\ref{fig:UI-Sync-W}.

%Moreover, while it is possible to automatically extract features that are indicative of UX problems (e.g., speech rate, tone, sentiment) from usability test audios as indicated in Section~\ref{explanations-content}, it is challenging to automatically detect the violations of Nielsen's heuristics, which is also part of the explanations.  
%Given the uncertainty in the automatic detection of UX problems and violations of heuristics, 

%We algorithmically extracted the speech rate, tone, and sentiment from the audio tracks of the usability test videos and used them as part of the explanations as indicated in Section~\ref{explanations-content}.
%With this hybrid WoZ design, we could control the quality of the AI-suggested problems and their explanations to better focus on answering the RQs.
\subsubsection{Wizard-of-Oz (WoZ) AI}
\label{sec:wozAI}

%\vera{to-do:  @Mingming @Vera add more discussions on the feasibility to implement such an AI and to generate explanation (R1). Perhaps using Fig 1 as an example, how to generate ``confused'' (R1).
%Also mention we will reflect on the limitations of WoZ and future work in discussions}

%\ming{@Vera: DONE. Please Check and update my revisions and comment out these comments.}

\rv{We introduce the details of our hybrid WoZ AI and discuss potential technical feasibility below. We will reflect on the limitations of WoZ and future work in Sec.~\ref{sec:limitations}.}

\label{WoZ_AI}
There are two pieces of information that the WoZ AI would offer to UX evaluators: \textit{AI-suggested problems} and \textit{AI explanations} to these problems. To generate AI-suggested problems, we first needed to find the \textit{ground-truth problems} in the usability test videos.

\textbf{Ground-truth Problems.} 
% \vera{double check the total number of ground truth problem, I think they are 23 and 10 instead of 20 and 8? also update the FP, FN manipulation}
To identify the ground-truth UX problems that users encountered in the two selected usability test videos, nine UX evaluators were recruited to independently review the videos, identify segments in the videos where users encountered problems, and write descriptions for the problems. 
Next, two other UX researchers reviewed the identified problems and their descriptions, discussed and consolidated a final list of ground-truth problems and descriptions. In total, there were 20 UX problems in the coffee machine video, and 8 in the website video.

\textbf{\rv{Supervised ML Predicting Usability Problems in Video Segments}} 
\rv{Our WoZ system was assumed to work with a supervised ML classifier trained on video data with labels of whether a segment contains a usability problem. Typically, the input features of a video-based supervised ML classifier include acoustic features (e.g., pitch) from the audio track, textual features from the transcription, and visual features from the video. To allow more human-consumable explanations, we also assume that more high-level semantic features, such as user actions and sentiments in the texts, can be obtained by either automatic recognition techniques or additional supervision}  

It is common sense that AI would hardly be perfect, especially in knowledge-rich domains, such as usability testing.
Thus, it would be inappropriate to have the WoZ AI suggest all the ground-truth problems.
To make the WoZ AI more realistic, we randomly added 5 (18\%) false problems (false positives) and removed 4 (14\%) true problems (false negatives) in the two selected videos in total.
%Specifically, we randomly added 20\% of the false problems and removed 20\% of the true problems.
%Thus, the overall accuracy of the WoZ AI was 0.8.
%We chose this performance for the WoZ AI as a recent paper suggests that end-users' expected accuracy of AI in general falls between .65 and .85~\cite{kocielnik2019will}.
%The corresponding explanations of the WoZ AI problems were used to generate the two textual features: "user said" and "user did."

\textbf{AI Explanations.} Our design of AI explanations is based on the potential technical feasibility of explainable AI, and informed by how \rv{usability} problems are identified and explained in UX practices~\cite{folstad2012analysis,fan2020Survey,mcdonald2012exploring}. Two forms of explanations are often generated by XAI techniques to explain an AI's decision for a particular instance~\cite{guidotti2018survey,adadi2018peeking,carvalho2019machine,gilpin2018explaining,arya2019one}: \textit{rule-based}---the decision rules that this instance violates or complies with; \textit{feature-based}---the features of this instance that are strong indicators for the decision. 
Correspondingly, we explain an AI-suggested problem in a usability test video segment by showing the \textit{UX design heuristics} that the video segment suggests the test product may violate, and the \textit{behavioral features} of the user that are indicative of the \rv{usability} problem. \rv{These two types of information about AI's predictions were also desired by UX evaluators when they worked with an AI agent to analyze usability test videos~\cite{fan2020vista,CoUX2021TVCG}}

\textit{UX Design Heuristics.}
UX literature has offered heuristics and principles for designing good user interfaces and identifying bad ones, such as Nielsen's heuristics~\cite{nielsen2005ten}, Norman's design principles~\cite{norman2013design}, and Gerhardt-Powals's cognitive engineering principles~\cite{gerhardt1996cognitive}. 
As these principles and heuristics are largely overlapped, we employed Nielsen's heuristics as they are commonly used in industry.
The original Nielsen's heuristics are abstract and require explanations to understand properly. 
As a result, we revamped the heuristics to make them more self-explainable while still keeping them short by referring to the explanations given by the Nielsen and Norman Group~\cite{nngroupheuristics}.
For example, the fourth heuristic is ``consistency and standards.''
The explanations of the violation of this heuristic is ``Using inconsistent words, visuals, or actions.''
%\vera{take a look at the sentence below}
To identify the violated heuristics for AI-suggested problems, we consulted the problem descriptions consolidated for the ground-truth problems and mapped them to the closest usability heuristics. 

\rv{There are several potential ways to implement such explanations. The heuristics themselves can be included as high-level features in the usability problem detection model and then directly provided as explanations. We can train additional supervised models to recognize these heuristics features. Recent research also began to investigate \mr{ways to} automatically detect the violations of usability heuristics. For example, Ponce et al.~\cite{ponce2018deep} proposed a convolution neural network model to detect the violation of three of the Nielsen's heuristics~\cite{nielsen2005ten}. Alternatively, one can leverage rationale or explanation generation techniques trained on human explanation data~\cite{ehsan2019automated,hind2019ted} to generate these design heuristics.}

\textit{Behavioral Features.}
The AI explanations also include the following behavioral features as indicators of the UX problems: \textit{speech rate}, \textit{tone (i.e., pitch)}, \textit{speech sentiment}, \textit{speech content}, \textit{what the user did in the video}. 
These features are chosen for two reasons. First, previous research suggests that UX evaluators should pay attention to what and how the user verbalizes and what the user does on the test product to identify UX problems ~\cite{mcdonald2012exploring,fan2020Survey}. For example, when encountering problems, users may slow down their speech, raise their tones, or use words with negative sentiments more often~\cite{fan2019concurrent, ericsson1984protocol,peterson1969concurrent,sporer2006paraverbal,kwon2003emotion}. Second, it is possible for current or near-future sensing and AI technologies to capture these behavioral features from usability test videos. 
\rv{For example, there is a rich body of literature and commercially available solutions for \textit{emotion} (e.g., confusion) and \textit{sentiment} analysis (e.g., iMotions~\cite{iMotions23}). Computer vision techniques, especially work on natural language video description~\cite{venugopalan2014translating,anne2017localizing}, can be used to identify and describe \textit{what the user did} in video segments.}

For WoZ AI, we adopted a hybrid approach to generate the behavioral features to explain the suggestion of a UX problem. The feature categories of \textit{speech rate}, \textit{tone}, and \textit{sentiment} were automatically extracted with the following algorithm.
%these types of features in a video and include them in a machine learning model for UX problem detection, for example through speech-to-text, sentiment analysis and video content analysis technologies.
%we designed the following algorithm to extract the speech rate, pitch, and sentiment features that reflect what and how users say. 
The audio track of each usability test video was divided into small segments based on silences in the user's think-aloud verbalizations and then transcribed into texts. 
For each segment, we calculated the speech rate by dividing the number of words spoken in the segment by its duration. 
%The number of words spoken in a segment was counted based on the text transcription of the segment. 
We calculated the tone (i.e., pitch) of the user by computing the fundamental frequency F0 (Hz) at the sampling rate of 100 Hz using the praatUtil library~\cite{pythonutil}.
For the sentiment, we analyzed the transcript using the VADER library~\cite{hutto2014vader} and then discretized it into three levels: negative, neutral, and positive.
%This hybrid WoZ AI was used in the four experimental conditions of our user study.
For the category of \textit{what user did}, we manually reviewed the videos and created a description. %Computer vision technique can be developed to generate such descriptions automatically in the future\cite{shao2020finegym}.

%\rv{As the computer vision community continues to push the boundary of human action understanding (e.g., \cite{shao2020finegym}), we believe automatic methods will soon be able to reliably understand ``what user did.''}\st{
\begin{figure}[!tb]
    \centering
    \includegraphics[width=\linewidth]{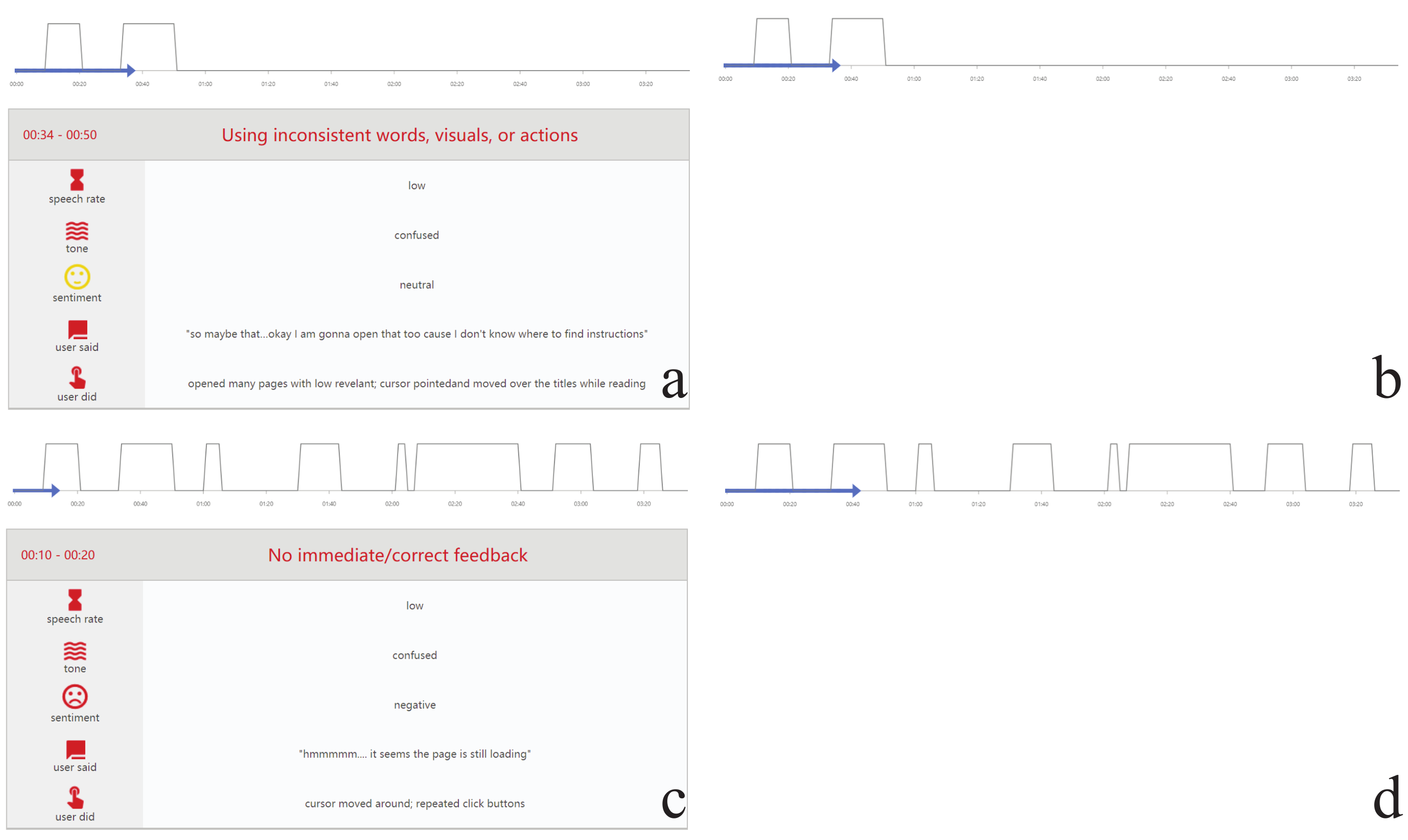}
    \caption{Comparison of the four different versions of AI Assistant (the bottom-left part of the user interface, as shown in Figure~\ref{fig:UI-Sync-W}de): (a) synchronously presenting the AI-suggested problems \textit{with} explanations (\textit{w/, sync}), (b) synchronously presenting the AI-suggested problems \textit{without} explanations (\textit{w/o, sync}), (c) asynchronously presenting the AI-suggested problems with explanations (\textit{w/, async}), and (d) asynchronously presenting the AI-suggested problems without explanations (\textit{w/o, async}), which is the baseline.}
    \label{fig:4UIs}
\end{figure}

\subsection{Design for Synchronization and Explanations}
\label{design-sync-exp}
As we were interested in studying the synchronization and explanations factors of AI, we designed AI Assistant with four alternative UIs to assist UX evaluators through an iterative design process.
The four versions include: (a) synchronously presenting the AI-suggested problems \textit{with} explanations (\textit{w/, sync}), (b) synchronously presenting the AI-suggested problems \textit{without} explanations (\textit{w/o, sync}), (c) asynchronously presenting the AI-suggested problems with explanations (\textit{w/, async}), and (d) asynchronously presenting the AI-suggested problems without explanations (\textit{w/o, async}).
Figure~\ref{fig:UI-Sync-W} shows the entire interface of the \textit{(w/, sync)} version of the AI Assistant.
Figure~\ref{fig:4UIs} shows the differences (in the bottom left part of Figure~\ref{fig:UI-Sync-W}de) among the four versions. The (\textit{w/o, async}) condition (Figure~\ref{fig:4UIs}d) is the \textit{baseline} since it is the simplest way of presenting UX problems among the four conditions.

%for the four experimental conditions: \textit{(sync, w)}, \textit{(sync, wo)}, \textit{(async, w)}, and \textit{(async, wo)}. 
\subsubsection{Design for Synchronization}
% \begin{figure}[tb]
%     \centering
%     \includegraphics[width=\linewidth]{Figures/figure2_revised.png}
%     \caption{(a) The Timeline for the \textit{async} versions of AI Aissitant, which reveals AI's suggestions of UX problems all at once. (b) The Timeline for the \textit{sync} versions, which reveals AI's suggestions of UX problems progressively up to current point a UX evaluator has viewed.}   
%     \label{fig:timelines}
% \end{figure}

The difference between the \textit{async} and the \textit{sync} versions of AI Assistant lies in how the AI-suggested problems are revealed to UX evaluators. 

In the \textit{async} way of presenting the AI, the AI-suggested problems are revealed to UX evaluators all at once at the beginning of their analysis and are always available during their analysis (Figure~\ref{fig:4UIs}cd). %(Figure~\ref{fig:timelines}a). 
This design simulates an \textit{asynchronous} workflow in which the AI completed its analysis before the UX evaluator. 
%UX evaluators would be able to see the AI's inferences for the whole video when they start their analysis.
A similar timeline was used in the literature~\cite{fan2020vista}. 

In the \textit{sync} way of presenting the AI, the AI-suggested problems are revealed to UX evaluators progressively as they play the video. 
This design simulates a \textit{synchronous} workflow in which the AI and the UX evaluator work through the video together at the same time (Figure~\ref{fig:4UIs}ab). %(Figure~\ref{fig:timelines}b).
UX evaluators are able to see the AI's suggestions up to the point where the video has been playing to, and cannot see the AI's suggestions for the rest of the video.
%Such UI design aims to create an impression that the AI is analyzing the video simultaneously with UX evaluators.
In cases where UX evaluators skip a portion of the video and start to play the video from a later timestamp, the AI's suggestions for the skipped portion are not revealed to them.  
But the AI's suggestions from that future timestamp onward will be progressively revealed to the UX evaluators as they continue playing the video.
We made this design choice to avoid UX evaluators from gaming the system by jumping to a very late timestamp of the video with the hope of revealing all the AI's suggestions. 

\subsubsection{Design for Explanations}
%\vera{@Mingming, per R1 comment, I also think this section can be cut down, not really necessary to show Fig 3}

%\ming{@Vera: DONE. Please Check and update my revisions and comment out these comments.}

\label{explanations-content}
%explanations provides UX evaluators with knowledge regarding why the AI thinks there is a problem. While to the best of our knowledge no current technologies provide explanations for UX problem detection algorithms, we consider what is possible from XAI literature~\cite{guidotti2018survey,adadi2018peeking,carvalho2019machine,gilpin2018explaining,arya2019one}, and what is desirable for UX evaluators to gain useful information. We followed an iterative design process to develop both the \textit{content} and the \textit{visual representation} of the explanations.

%\textbf{Content.} 
For the \textit{w/} problem explanations versions of AI Assistant, the AI explanations Panel (Figure~\ref{fig:4UIs}ac) %(Figure~\ref{fig:UI-Sync-W}e) 
contains two types of information: the UX design heuristics that are violated and the behavioral features that suggest UX problems, as described in Section~\ref{WoZ_AI}.
The key design consideration for presenting the explanations content is to ensure it is \textit{light-weight} and \textit{scannable}.  
UX evaluators need to attend to different types of information---the video to be analyzed and the AI-suggested problems---while writing their own problem descriptions using the annotation function.
Thus, the visual representation of the explanations should minimize the attention needed and be scannable without extensive reading. 

Following the design consideration, we minimized the amount of text by using visual icons that conveyed information effectively and organized the features in a consistent and scannable format. 

\subsection{Implementation}
\label{implementation}
%We implemented the AI-assisted tool with four UIs for the four experimental conditions as a web application that can be assessed remotely using a web browser. 
%The tool includes a front-end UI that is built on top of React and Mobx libraries and a back-end Nodejs server. 
%We deployed the tool on Amazon Web service (AWS) to allow participants to access to the tool remotely and robustly.

We implemented the four versions of AI Assistant %corresponding to the four conditions of the two factors (i.e., synchronization and explanations) 
as a web application, which would allow for recruiting a broad set of UX evaluators to participate in the study.  
We deployed the tool on Google Firebase and the videos on Amazon Web service (AWS) to allow UX evaluators to access the tool remotely and robustly.

The tool includes a back-end NodeJS server and a front-end UI that is built on top of React and Mobx libraries. 
The NodeJS server provides an application programming interface (API) to serve the links of the usability test videos from Amazon S3 to our front-end UI seamlessly. 
The AI-suggested problems and the corresponding explanations for each usability test video are stored into a JSON file, which is visualized on the front-end UI in real-time depending on the experimental condition. 
The front-end UI has been revamped by utilizing React Router into the application, which allows us to control the experimental conditions that UX evaluators have access to by sharing them with the corresponding URLs. 
By using Mobx, the activity history of the video player (e.g., how the video is played, paused, rewind) and the usability problems identified by UX evaluators are stored in the state during the user study and saved into the JSON format log data, which will be analyzed as part of the quantitative data of the study.

%% file: texfiles/4-userstudy.tex
\section{User Study}
To answer the RQs, we used a mixed-method, which included a controlled user study, questionnaires, and interviews. In the controlled study, UX evaluators analyzed two usability test videos with AI Assistant and answered questionnaires regarding their use of AI Assistant. 
Post-study interviews offered more insights into their preferences of AI Assistant.

\subsection{Participants}
We recruited participants from industry and local universities via electronic flyers with two inclusion criteria: having at least one year's experience in UX and having experience of analyzing think-aloud usability test sessions. 
%We also encouraged participants to contact their friends or colleagues who might also be interested in the study. 
We first conducted a pilot study with two participants, who had 1-2 years of UX/HCI experience. 
The two pilot study sessions helped us to adjust the whole study procedure and duration, ensuring that the study design was appropriate.

We then recruited 24 participants as UX evaluators for the actual user study. 
%The demographic information is shown in Table 1. 
Among the 24 participants, sixteen were females and eight were males; thirteen were at the age of 25-34 and eleven at the age of 18-24.
In terms of educational backgrounds, fifteen had Bachelor's degrees, four had Master's degrees, two had Doctoral degrees, and three had some college degrees.
Regarding the UX/HCI work experience, fourteen participants had 1-2 years of experience, six had 2-3 years of experience, and four had more than 3 years of experience. 

\begin{table}[tb]
\newcommand{\sync}{\textcolor{teal}{sync}}
\newcommand{\async}{\textcolor{violet}{async}}
\newcommand{\w}{\textcolor{black}{w/}}
\newcommand{\wo}{\textcolor{gray}{w/o}}
\newcommand{\dvid}{\textcolor{orange}{video-1}}
\newcommand{\wvid}{\textcolor{olive}{video-2}}

\caption{Participant information and the study design. For trust in AI, higher score indicates more trust; for experimental conditions, video-1 is the physical device video and video-2 is the digital website video.} 
\label{tab:participants}
\resizebox{\linewidth}{!}{
\small
\begin{tabular}{lm{3.2cm}m{2.2cm}m{2cm}m{6cm}}
\toprule
% \textbf{ID} & \textbf{Knowledge about Nielsen's Heuristics} & \textbf{Knowledge about AI} & \textbf{Trust in AI} & \textbf{Experimental conditions} \\
%  & A little–have learned some of them; & No knowledge & the higher the score, & video-1: the physical device video;\\
%   & Some-have learned all of them; & A little-know very basic concepts &  the higher the trust & video-2: the digital website video\\
%     & A lot-have frequently applied them in work; & Some-have used AI/ML & & \\
\textbf{ID} & \textbf{Knowledge about Nielsen's UX design Heuristics~\cite{nielsen2005ten}} & \textbf{Knowledge about AI} & \textbf{Trust in AI} & \textbf{Experimental conditions} \\
\midrule
P1 & Some
& A little
& 4 & (\w,\async, \dvid), (\w, \sync, \wvid)
\\
P2 & Some
& A little
& 4 & (\wo, \sync,  \dvid), (\wo, \async, \wvid)
\\
P3 & A lot
& A little
& 4 & (\wo, \async, \dvid), (\wo, \sync, \wvid)
\\
P4 & Some
& A little
& 5 & (\w, \sync, \wvid), (\w, \async, \dvid)
\\
P5 & A lot
& A little
& 5 & (\w, \async, \wvid), (\w, \sync, \dvid)
\\
P6 & A lot
& A little
& 4 & (\wo, \sync, \wvid), (\wo, \async, \dvid)
\\
P7 & Some
& No knowledge
& 4 & (\wo, \async, \wvid), (\wo, \sync, \dvid)
\\
P8 & Some
& A little
& 3 & (\w, \sync, \dvid), (\w, \async, \wvid)
\\
P9 & A lot
& A little
& 4 & ( \w, \async, \dvid), (\w, \sync, \wvid)
\\
P10 & Some
& A little
& 3 & (\wo, \sync, \dvid), (\wo, \async, \wvid)
\\
P11 & Some
& No knowledge
& 2 & (\wo, \async, \dvid), (\wo, \sync, \wvid)
\\
P12 & A lot
& Some
& 4 & (\w, \sync, \wvid), (\w, \async, \dvid)
\\
P13 & A lot
& A little
& 3 & (\w, \async, \wvid), (\w, \sync, \dvid)
\\
P14 & Some
& Some
& 3 & (\wo, \sync, \wvid), (\wo, \async, \dvid)
\\
P15 & Some
& A little
& 4 & (\wo, \async, \wvid), (\wo, \sync, \dvid)
\\
P16 & Some
& Some
& 4 & (\w, \sync, \dvid), (\w, \async, \wvid)
\\
P17 & A lot
& Some
& 3 & (\w, \sync, \dvid), (\w, \async, \wvid)
\\
P18 & Some
& A little
& 4 & (\w, \async, \dvid), (\w, \sync, \wvid)
\\
P19 & A lot
& A little
& 4 & (\wo, \sync, \dvid), (\wo, \async, \wvid)
\\
P20 & A lot
& A little
& 3 & (\wo, \async, \dvid), (\wo, \sync, \wvid)
\\
P21 & A little
& Some
& 2 & (\w, \sync, \wvid), (\w, \async, \dvid)
\\
P22 & Some
& A little
& 3 & (\w, \async, \wvid), (\w, \sync, \dvid)
\\
P23 & A lot
& A little
& 3 & (\wo, \sync, \wvid), (\wo, \async, \dvid)
\\
P24 & A lot
& A little
& 4 & (\wo, \async, \wvid), (\wo, \sync, \dvid)
\\
\bottomrule
\end{tabular}
}
\end{table}

\subsection{Experimental Design}
% \ming{TODO: mention (async, w/o) is the baseline}
% The \textit{synchronization} factor refers to the time when the AI's inferences are revealed to UX evaluators. 
% We follow two common ways in which two human evaluators collaborate with each other: \textit{synchronously} and \textit{asynchronously}. 
% Similarly, the AI could also reveal its inferences \textit{synchronously} (i.e., sync)  and \textit{asynchronously} (i.e., async). 
% For the \textit{explanations} factor, we present the UX evaluators with the AI that is accompanied \textit{with} and \textit{without} explanations of how the AI works.  
We employed a 2-by-2 mixed design, with the \textit{explanations} as the between-subjects factor and the \textit{synchronization} as the within-subject factor. This means each participant was randomly assigned to use AI Assistant either \textit{w/} explanations or \textit{w/o} and to complete two usability video analysis tasks, one of which was with the \textit{sync} AI Assistant and the other was with the \textit{async}. We counter-balanced the order of the \textit{sync} and \textit{async} AI Assistants and randomized the order of the two videos. 

We chose this mixed design for the following reasons. First, there is a potential learning effect for explanations. After participants see the explanations for the first test video, they might be primed to use similar information when analyzing the second video even if they are not given explanations. Thus, we set the explanations as the between-subjects factor. Second, there is no foreseen learning effect for synchronization so it can be set as the within-subject factor. Doing so also allows each participant to compare their experiences with \textit{sync} and \textit{async} human-AI collaboration. %Moreover, this mixed-design allowed us to recruit relatively fewer UX evaluators than a fully between-subjects design.

%We counter-balanced the videos and the tool's UI versions 
%according to the last column of the Table ~\ref{tab:participants} shows  
%so that each participant analyzed one video using a \textit{sync} AI and the other video using a \textit{async} AI, both \textit{w} or \textit{w/o} explanations.
% This experimental design choice, which was determined after two pilot studies with two UX evaluators, kept the study length manageable and allowed every participant to experience and compare \textit{sync} and \textit{async} AI Assistant, which we inquired in the qualitative study.
%We decided not to set the two factors both as within-subject factors to avoid overloading the study participants and to keep the study length manageable. 
%Our pilot studies with two UX evaluators found that the mixed study design took them over one hour to complete, in which each one analyzed two different videos.
%Thus, with two within-subject factors, each participant would have to analyze four different videos and the whole study session could have been over two hours. 
% \jian{Still not justify why we use sync to be the within factor, but not explanations? Can we say sync/async hasn't been studied before, so we pro this?}

%there are four experimental conditions in which the AI presents its suggested problems and explanations to UX evaluators: sync with explanations \textit{(sync, w/)}, sync without explanations \textit{(sync, w/o)}, async with explanations \textit{(async, w/)}, and async without explanations \textit{(async, w/o)}.

\subsection{Procedure}
%We deployed the AI Assistant as a web application, as described in Section~\ref{implementation}, to allow for remote participation as the COVID pandemic prevented in-person studies. 
The study was divided into the following phases: \textit{set-up}, \textit{pre-test questionnaire}, \textit{training session}, \textit{two formal-task sessions} (each including one formal task and one post-task questionnaire), and \textit{post-test questionnaire and interview}. All studies, including the pilot studies, were conducted online. The study took 75 minutes on average to finish, and each participant was compensated with \$15. 

\textbf{Set-up.} Participants were asked to enter our ZOOM room, check their microphone, speaker and camera, and to share their screen with the moderator.

\textbf{Pre-test Questionnaire}. The moderator sent participants the URL to the online questionnaire via ZOOM chat.
In the pre-test questionnaire, participants were asked to fill in their basic demographic questions (e.g., age, gender, education backgrounds), years of UX/HCI work experience, and their experience of conducting usability testing sessions.
Since they would analyze the usability test videos with the AI, they were also asked about their knowledge of AI and ML (four levels from no knowledge to a lot of knowledge). 
Moreover, they were also asked to rate their general trust in AI and ML on a 5-point Likert scale. 
%Table~\ref{tab:participants} shows the answers from the pre-test questionnaire.

\textbf{Training Session.} We asked participants to learn and practice using the AI assistant to analyze a usability test video different from the ones used in the formal task phase. 
We first introduced to participants how to play the video and write problem descriptions.
We then explained the visualization of the AI's inferences. 
For the participants in \textit{(w/, sync)} and \textit{(w/, async)} conditions that showed the AI's explanations, we explained to them that the explanations consisted of 1) the UX design heuristics that were violated; and 2) the input features that the AI considered when making its predictions. We kept the introduction consistent for all participants in these conditions.
We informed participants that they could decide whether to use the AI or not. 
We then showed them with an example of how to find and record a usability problem. 
Next, we asked participants to practice using the tool to find and record one usability problem on the practicing video.

\textbf{Formal Task Sessions.} Participants used two versions of AI Assistant to review the two usability test videos (described in Section~\ref{usability-test-videos}) to identify problems and write problem descriptions. As discussed in the previous section, one task was with the \textit{sync} AI Assistant and the other with the \textit{async} AI Assistant, and the order of the two versions of AI Assistant was counter-balanced.
For each usability test video, the moderator suggested (but not enforced) participants spend no more than about two times of the video length to keep the study on time. 
After completing each task, the participants were asked to complete a post-task questionnaire. 
%During the entire study, the moderator closely monitored the participants to ensure participants'.

\textbf{Post-test Questionnaire and Interview.} After finishing reviewing each of the two usability test videos, the participants were asked to complete a post-task questionnaire reporting their satisfaction, understanding, and trust in AI Assistant (details in Section~\ref{measure}). After completing both tasks, we conducted semi-structured interviews for 10-20 minutes regarding participants' experiences (e.g., how did you use AI Assistant? what did you like and dislike about AI Assistant? why?) and preferences (e.g., which version of AI Assistant did you prefer more? what other features of AI Assistant would you want? why?) of using AI Assistant, as well as other customized questions regarding the moderator's observations of their behaviors in the study.

%\vera{you may want to add a little more details about the interview since it is a mixed-methods study, is it semi-structured? what kind of questions? how long did it last?}
%Participants were asked to complete a post-test questionnaire and to be interviewed regarding their experiences and preferences of using the two different AI-Assistants in the study. 

\subsection{Measurements}\label{measure}
% Survey: understanding, trust, satisfaction, attribution; (control) AI knowledge, AI trust, UX experience

% Task performance measure: problem found, problem precision, recall, time, problem description content and quality

% Behavioral metrics: pause, pause time, jump, rewind
% \jian{Below is my writing for this subsection.}

From the experiment, we captured three types of measurements, including task performance metrics related to identified UX problems, behavioral metrics reflecting interaction patterns, and subjective perception of AI Assistant measured by survey responses.

\textbf{Task Performance.} This type of measures focused on the number of UX problems found by each participant. Given the ground truth of UX problems in the two usability tasks and the problems identified by a participant, we calculated the \textit{precision}---the percentage of correct problems among all problems identified by the participant, and \textit{recall}---the percentage of correct problems identified by the participant among all correct problems existed in the ground truth. To further verify the effect of AI Assistant on participants' task performance, we also analyzed the \textit{overlap} between a participant's identified problems and AI Assistant's suggestions. We also looked at the content of \textit{problem descriptions} written by participants as a secondary measurement of task performance. 

%For the task performances, we calculated the task completion time, the number of UX problems found, and the precision and recall of the found problems. \jian{Do we need to explain the precision and recall here?} We also analyzed the content and quality of the problem descriptions generated by evaluators. Specifically, we computed the number of overlapping AI-inferred problems with an evaluator-identified one, the length of problem descriptions, and the number of overlapping words between their descriptions and the corresponding ones provided by AI. 

\textbf{Behavioral Metrics.} These were calculated from the tool's log data, including the total \textit{time} spent on an analysis, the number of \textit{pauses}, the duration of \textit{pause time}, the number of forward \textit{jumps}, and the number of \textit{rewinds}.
We considered a pause when a participant paused the video for 3 seconds or more, to avoid counting any unintentional actions. A forward jump was defined as a participant fast-forwarded the video by clicking on the timeline with more than 2 seconds apart from the current video time. Similarly, a rewind was counted when a participant went back in the video time by 2 seconds or more. 

\textbf{Subjective Perception.} We relied on questionnaire responses to measure participants' \textit{satisfaction}, \textit{understanding}, and \textit{trust} of AI Assistant. Specifically, satisfaction was measured by a three-item scale based on the After-Scenario Questionnaire~\cite{lewis1995computer} (Cronbach's alpha $= 0.72$). Understanding was measured by two self-reported items: ``I felt that I had a good understanding of how AI Assistant works'' and ``I felt that I had a good understanding of why AI Assistant detects UX problems'' (Cronbach's alpha $= 0.86$). Trust was measured by a three-item scale (e.g. ``I feel like I can count on AI Assistant to provide reliable suggestions for analysis of think-aloud sessions'') adapted from ``trust intention'' in McKnight’s framework on Trust~\cite{mcknight1998initial,mcknight2002developing} (Cronbach's alpha $= 0.72$). All items were rated on a 7-point Likert scale. 
The Cronbach's alpha values, as indicated above, showed high internal consistency in the questionnaire responses. 
 
% , UX evaluators answered questions on a 1-7 Likert scale regarding their understanding, trust, satisfaction, and attribution of the AI-assistants. Their preferences with the different AI-assistants (i.e., synchronous or asynchronous) were also recorded, along with their willingness of using the tool(s).

\subsection{Analysis Methods}\label{analysis}
For quantitative data, we computed descriptive statistics of the corresponding measures in Section~\ref{measure} and performed mixed-effects regression models, which will be described in detail in Section~\ref{quantitative-results}.
% \ming{@Vera: add details about the mixed-effect model.}\vera{It is in 5.1, I think the new 2AC missed it...is it better to move here? }
For qualitative data, the interview recordings were first transcribed into texts, and two researchers of the team performed thematic analysis on the texts independently and discussed the common themes that emerged from the texts. Finally, the themes were further discussed with two additional researchers of the team and consolidated into the key findings, which will be described in detail in Section~\ref{qualitative_results}.  

%% file: texfiles/5-quantitativeresults.tex
\section{Results}
We first present quantitative results from the experiment and then qualitative results from the interviews. We also annotate each subsection regarding the RQs that they primarily answer.

%\jian{add an overview paragraph? also remind the mixed methods?}

\subsection{Quantitative Results}

%\vera{to-do:@Vera re-write the wording in response to R2}
\label{quantitative-results}

\textbf{Statistical method overview.} We conducted quantitative analysis on various dependant variables related to participants' task performance, behavioral patterns interacting with AI Assistant and subjective perception of the tool. We normalized the measurements that were likely impacted by either the video length or the number of ground-truth UX problems, detailed in each analysis below. For each dependant variable,  we performed a separate mixed-effects regression with \textit{explanations} (\textit{w/} or \textit{w/o}) and \textit{synchronization} (\textit{sync} or \textit{async}) as the fixed effects and \textit{participants} as the random effects. The regression model also included the interaction effect between explanations and synchronization. The regression model further included the \textit{usability test videos} (website or coffee machine), participants' self-reported \textit{UX experience}, \textit{knowledge of AI} and general \textit{trust in AI}, \textit{age group}, and \textit{gender} as control variables. All tests were performed with the \textit{nlme} package in R. For each test, we checked for outliers with the dependant variable as outside 1.5 times the interquartile range \rv{(we will only mention below if a test had outliers identified and removed)}, and made sure there was no multicollinearity (all $VIF<3$). 
Below we report the descriptive statistics (mean values and standard division) for each of the quantitative measures and significant results from the regression analysis. 

Technical breakdowns happened for three participants: both tasks for P2 (\textit{w/o explanations}) and P12 (\textit{w/ explanations}), and one task (\textit{async}) for P9 (\textit{w/ explanations}), so we removed these five data points. On average, for the website video, participants spent 643 seconds (SD=48.5) and found 4.86 (SD=0.45) UX problems (ground-truth UX problems is 8; video length is 3.57 minutes). For the coffee machine video, participants spent 1331 seconds (SD=76.9) and found 11.9 (SD=1.04) UX problems (ground-truth UX problems is 20; video length is 11.37 minutes).
%\st{We normalized the measurements that were likely impacted by either the video length or the number of ground-truth UX problems.}

\subsubsection{Task Performance (RQ1, RQ2)}
\label{taskperformance}

\textbf{Overall performance measurements.} We started by analyzing the measurements reflecting how well participants performed the tasks as described in Section~\ref{measure}, specifically the total numbers of UX problems identified,  the Precision and Recall of their identified problems based on the ground truth problems. Given the difference between the two usability tasks (20 UX problems in the coffee machine video, 8 in the website video), we normalized the total number of UX problems found by the number of ground-truth problems in each video. Descriptive statistics, including means and standard deviations, are presented in Table~\ref{tab:task} (Columns N Problems, Precision, and Recall).

% \vera{to-do: @Mingming R1 wants to see \% of accurate AI suggestions confirmed, it needs calculation from column 1 and 2 below. Should we add an additional column or respond?}

% \ming{@Vera: Yes, I agree. Could you calculate \% of accurate AI suggestions confirmed and add it to Table 3?}

% \vera{can you help?}

\begin{table}
\centering
  \caption{The mean and standard deviation (in parenthesis) of task performance measures: the number of UX problems identified by participants (N Problems) normalized by the number of ground-truth problems in each video, the precision and recall of the identified problems, and the length of problem description for each entry of UX problem (Desc. Len.)}\label{tab:task}
 {\footnotesize
  \begin{tabular}{ccp{1.7cm}p{1.7cm}p{1.7cm}p{1.7cm}}
%   \begin{tabular}{Cc^c^c^c^c^c^c^c^c^c}
  
    \toprule
    \rowstyle{\bfseries}
         &  & \textbf{N Problems (normalized)}   &\textbf{Precision} & \textbf{Recall} &  \textbf{Desc. Len.} \\
    \midrule
    & \textit{async} &0.69 &.860 &.560 &11.4 \\
    \textit{w/}&&(0.09) &(.049)&(.055) &(2.2)\\
    & \textit{sync} &0.56 &0.865 &.475 &10.2 \\
    &&(0.06) &(.039)&(.047) &(1.9) \\

    \midrule
    &\textit{async} &0.50 &.939 &.448  &11.3 \\
    \textit{w/o}&&(0.07) &(.037)&(.047)&(1.8) \\
    &\textit{sync} &0.66   &.894 &.589 &13.5 \\
    &&(0.08) &(.051)&(.068) &(2.1) \\

  \bottomrule
\end{tabular}
}
\end{table}

We performed a mixed-effects regression model, as described above, on the number of identified problems (normalized by number of ground-truth problems), Precision, and Recall respectively. 
For the number of identified problems (normalized), we found a significant two-way interaction between the presence of explanations and synchronization ($\beta=-0.29, SE=0.13,F(2,18)=4.54, p<0.05$). \rv{Post-hoc analysis using \textit{emmeans} package of R found the contrast between \textit{async} and \textit{sync} marginally significant for AI Assistant \textit{w/o explanations} ($p=0.10$)\footnote{Given the relatively small sample size, we consider $p<0.05$ as significant, and $0.05 \leq p<0.10$ as marginally significant, following statistical convention~\cite{cramer2004sage}}, but not significant for AI Assistant \textit{w/ explanations} ($p=0.22$), suggesting that the interactive effect was mainly caused by the difference of synchronization made for participants interacting with AI Assistant \textit{w/o explanations}.}

\begin{figure*}[ht]
    \centering
\begin{subfigure}[b]{.47\textwidth}
  \centering
  \includegraphics[width=\linewidth]{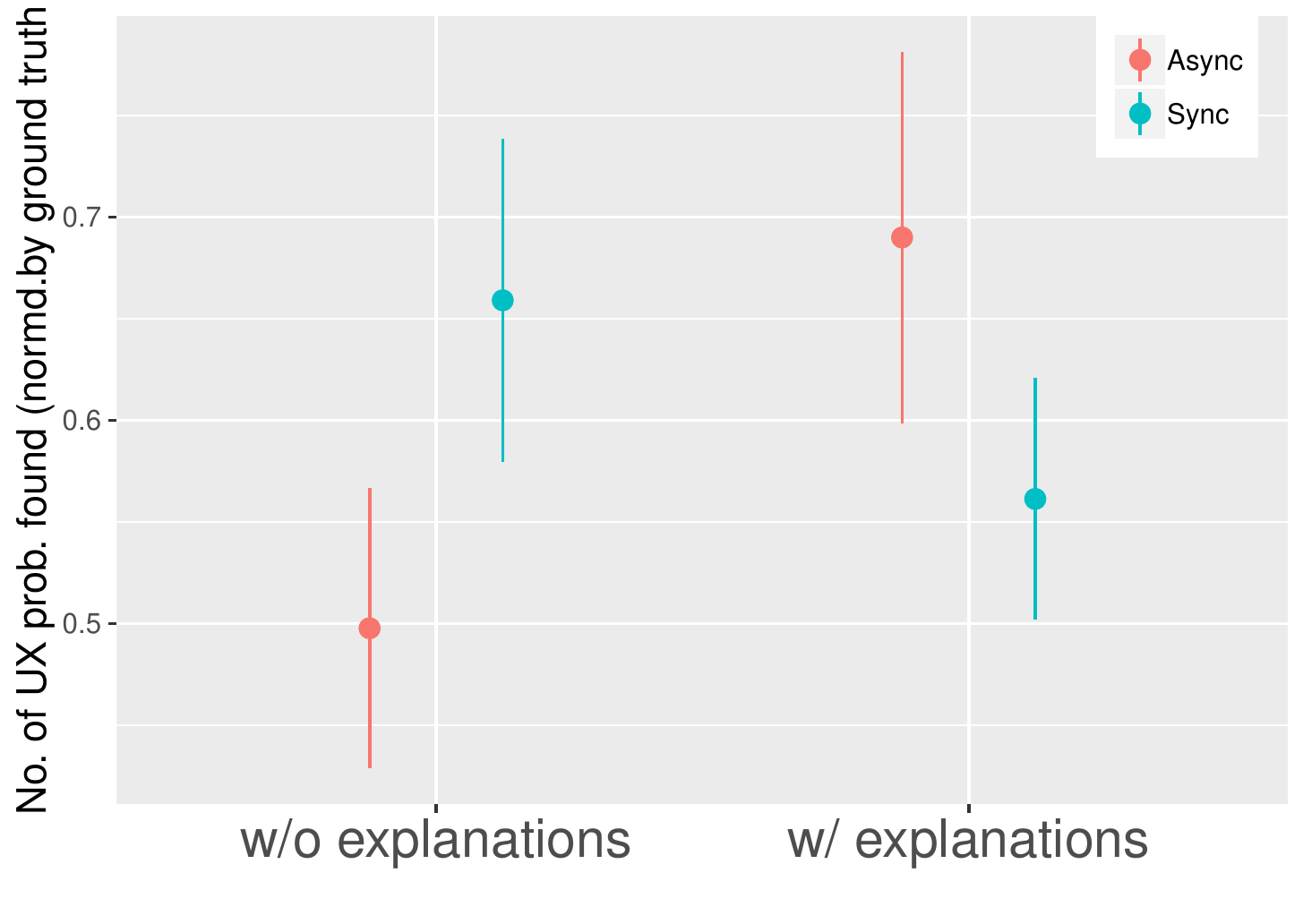}
 % \caption{Customer profile presented in all conditions for annotation}
 % \label{fig:interface_1}
 % \label{fig:sub1}
\end{subfigure} %
% \begin{subfigure}[b]{.47\textwidth}
%   \centering
%   \includegraphics[width=\linewidth]{Figures/precision.pdf}
%   %\caption{explanations and questions presented in the XAL condition}
%   % The final caption as appearing in the pdf needs reformating. Instead of having two different captions, you can have a single caption where you can manually specify (a) and (b). 
% \end{subfigure}
% \begin{subfigure}[b]{.47\textwidth}
%   \centering
%   \includegraphics[width=\linewidth]{Figures/recall.pdf}
%  % \caption{Customer profile presented in all conditions for annotation}
%  % \label{fig:interface_1}
%  % \label{fig:sub1}
% \end{subfigure} %
\begin{subfigure}[b]{.47\textwidth}
  \centering
  \includegraphics[width=\linewidth]{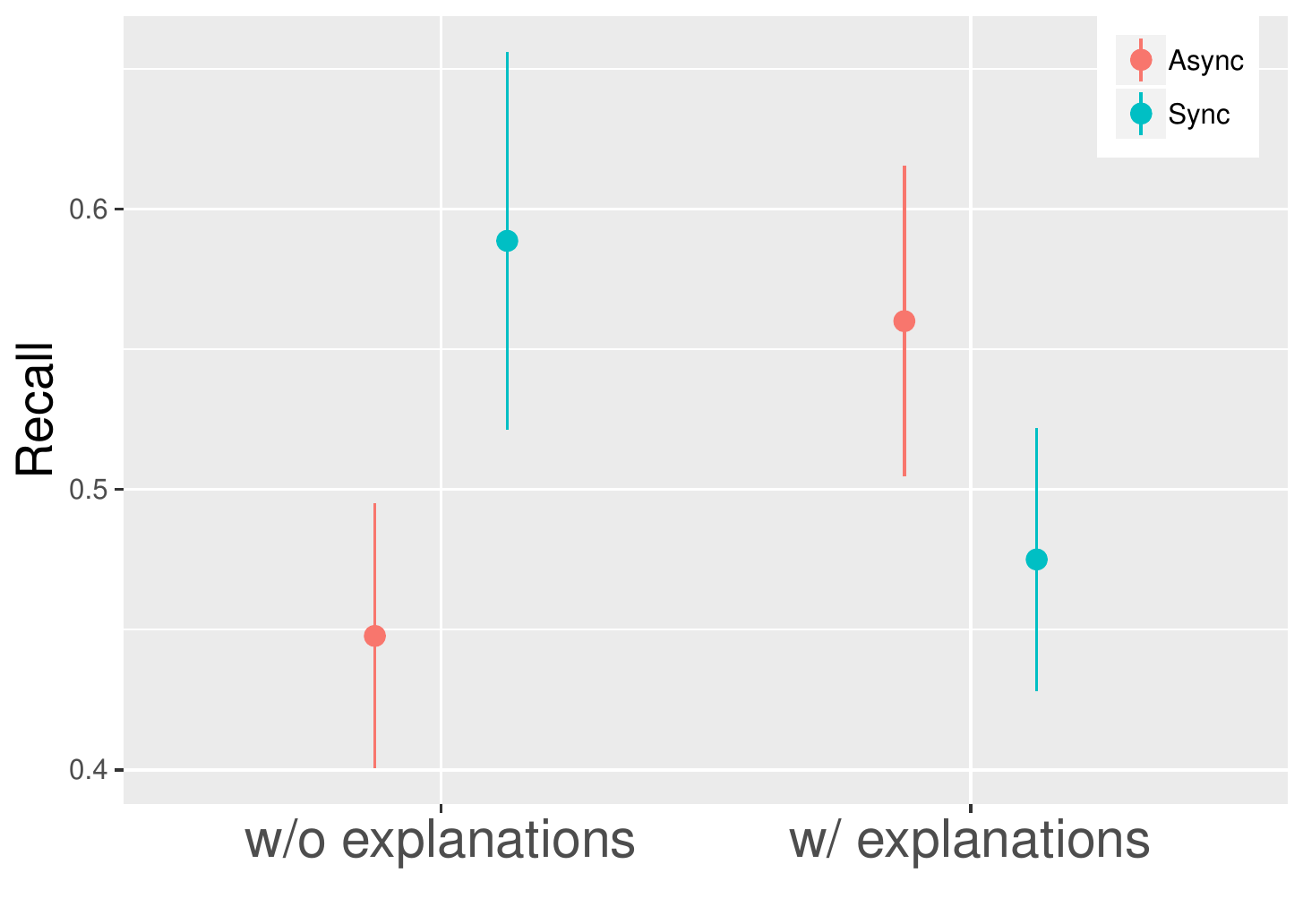}
  %\caption{explanations and questions presented in the XAL condition}
  % The final caption as appearing in the pdf needs reformating. Instead of having two different captions, you can have a single caption where you can manually specify (a) and (b). 
\end{subfigure}
    % \caption{Number of UX problems identified, precision, recall, and found UX problems overlapping with AI suggestions by participants across conditions. All error bars represent +/- one standard error.}
        \caption{Two-way interaction between explanations and synchronization on the number of UX problems found (normalized by the number of problems in ground truth), and Recall, as shown by means and standard divisions across conditions. All error bars represent +/- one standard error. }
    \label{fig:problem}
\end{figure*}

While we did not find any significant effect on Precision, we found the same significant two-way interaction between the presence of explanations and synchronization on Recall ($\beta=-0.23, SE=0.09,F(2,18)=6.52, p=0.02$\rv{; post-hoc analysis found the contrast between \textit{async} and \textit{sync} significant for AI \textit{w/o explanations} ($p=0.04$), but not significant for AI \textit{w/ explanations} ($p=0.17$)}). No main effect was found. These two-way interactions on total number of problems and Recall are illustrated in Figure~\ref{fig:problem}: when interacting with AI Assistant \textbf{\textit{w/o explanations}}, participants found significantly \textbf{more UX problems} from the ground truth, in the \textbf{\textit{sync}} than the \textbf{\textit{async}} condition. For those interacting with AI Assistant \textbf{\textit{w/ explanations}}, they did not show such difference in the \textit{sync} than the \textit{async} condition.
%\st{, but they generally found \textbf{more UX problems} than the \textbf{baseline} condition (\textit{w/o, async})}.
%We tested the main effect of synchronization on the number of identified problems for participants used AI Assistant w/ and w/o explanations separately. We found a positive main effect of synchronization for participants interacted with AI Assistant without explanations ($\beta=0.072, SE=0.033,F(2, 18)=2.13, p<0.05$), but no significant effect for those with explanations. 
%It confirmed that the main reason for the interaction effect is the difference of synchronization made for the AI Assistant without explanations.

\begin{table}[tbh!]
\centering
  \caption{The mean and standard deviation (in parenthesis) of measurements on overlaps with AI's suggestions and mistakes: the percentage of identified UX problems overlapped with AI's suggestions over the total number of AI suggested problems (Pct. AI suggestions Confirmed), 
  overlap percentages over AI's true positive suggestions (Pct. AI TP Confirmed), 
  overlap percentages over AI's false positive suggestions (Pct. AI FP Confirmed), and overlap percentages over AI's false negative suggestions (Pct. AI FN Found).\st{, and the percentage of words in a  problem description for each entry of UX problem overlapped with the AI's explanations (Pct. Desc. Overlap).}}\label{tab:overlap}
 {\footnotesize
  \begin{tabular}{ccp{2.4cm}p{1.8cm}p{1.8cm}p{1.8cm}}
%   \begin{tabular}{Cc^c^c^c^c^c^c^c^c^c}
  
    \toprule
    \rowstyle{\bfseries}
         &  & \textbf{Pct. AI suggestions Confirmed} & \rv{\textbf{Pct. AI TP Confirmed}} & \textbf{Pct. AI FP Confirmed} & \textbf{Pct. AI FN Found} \\
    \midrule
    & \textit{async} &53.2\% & 55.4\% & 21.7\%  & 23.3\%  \\
    \textit{w/}&&(6.3\%) & (20.8\%) & (9.3\%) &(10.0\%) \\
    & \textit{sync} &47.5\% & 60.5\% & 19.7\% & 12.1\%  \\
    &&(4.7\%) & (20.1\%) & (7.0\%) &  (5.1\%) \\

    \midrule
    &\textit{async}  &40.6\% & 52.6\% & 3.0\%  & 12.1\% \\
    \textit{w/o}& &(4.2\%) &  (19.7\%) & (3.0\%)&(9.3\%) \\
    &\textit{sync}&56.5\% &  62.0\% & 19.7\%  & 21.2\%  \\
    & &(6.0\%)  & (24.9\%) & (7.4\%) & (10.3\%)  \\

  \bottomrule
\end{tabular}
}
\end{table}

\textbf{Overlaps with AI.}
We further examined how the AI influenced participants' analysis by the percentage of their identified UX problems overlapped with AI's suggestions, as divided by the total number of AI suggested problems (Pct. AI suggestions Confirmed), as well as overlap percentages over AI's false positive suggestions (Pct. AI FP Confirmed) and false negative suggestions (Pct. AI FN Found). Specifically, Pct. AI suggestions Confirmed indicates participants' propensity to confirm AI's suggestions. Pct. AI FP Confirmed indicates participants' tendency to be misled by AI's suggestions when there were no UX problems; and Pct. AI FN Found indicates participants' ability to identify a correct UX problem when AI failed to make a suggestion. Descriptive statistics of these measurements are presented in Table~\ref{tab:overlap}. 

\begin{figure*}[tbh!]
    \centering
\begin{subfigure}[tbh!]{.47\textwidth}
  \centering
  \includegraphics[width=\linewidth]{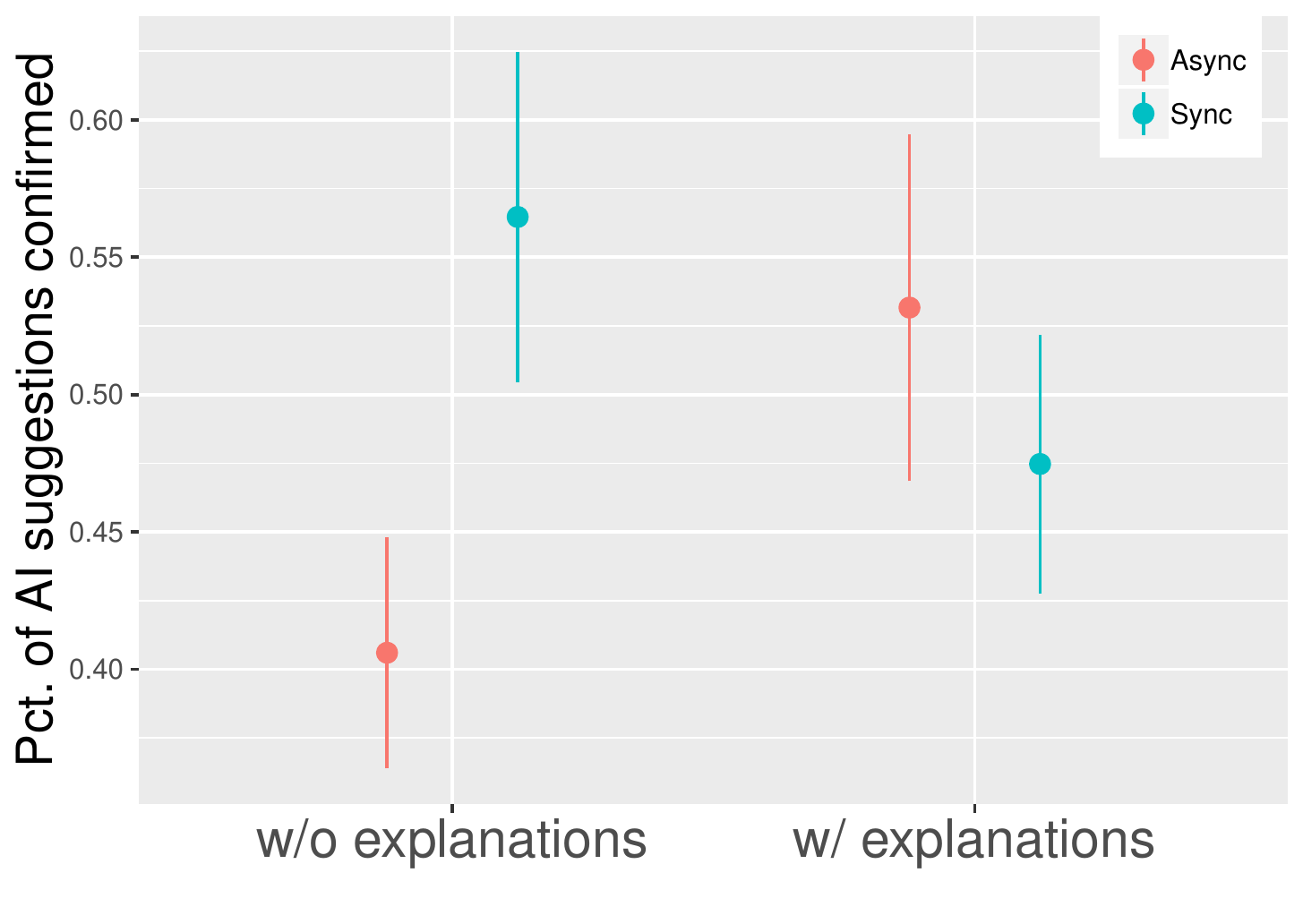}
 % \caption{Customer profile presented in all conditions for annotation}
 % \label{fig:interface_1}
 % \label{fig:sub1}
\end{subfigure} %
\begin{subfigure}[tbh!]{.47\textwidth}
  \centering
  \includegraphics[width=\linewidth]{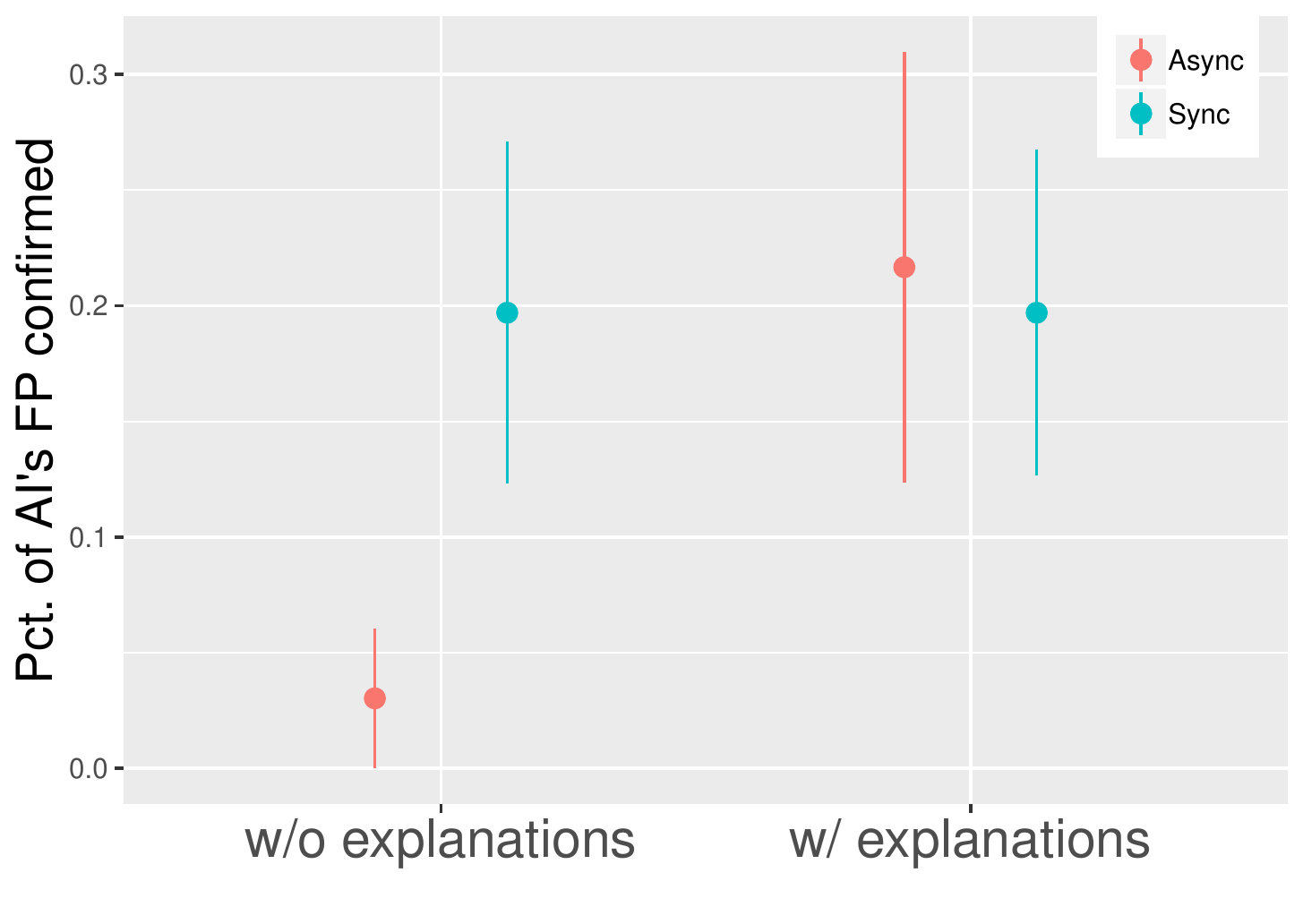}
  %\caption{explanations and questions presented in the XAL condition}
  % The final caption as appearing in the pdf needs reformating. Instead of having two different captions, you can have a single caption where you can manually specify (a) and (b). 
\end{subfigure}
    % \caption{Number of UX problems identified, precision, recall, and found UX problems overlapping with AI suggestions by participants across conditions. All error bars represent +/- one standard error.}
        \caption{Two-way interaction between explanations and synchronization on the percentage of AI suggestions confirmed, and percentage of AI's false negative suggestions confirmed, as shown by means and standard divisions across conditions. All error bars represent +/- one standard error. }
    \label{fig:overlap}
\end{figure*}

We performed separate mixed-effects regression analysis as described earlier on each of the three measurements. We found the two-way interaction between the presence of explanations and synchronization to be significant for Pct. AI suggestions Confirmed ($\beta=-0.22, SE=0.09,F(2,18)=6.94, p=0.02$ \rv{; post-hoc analysis found the contrast between \textit{async} and \textit{sync} significant for AI \textit{w/o explanations} ($p=0.02$), but not significant for AI \textit{w/ explanations} ($p=0.26$)}) and marginally significant for Pct. AI FP Confirmed ($\beta=-0.18, SE=0.10,F(2,18)=3.43, p=0.08$\rv{; post-hoc analysis found the contrast between \textit{async} and \textit{sync} significant for AI \textit{w/o explanations} ($p=0.03$), but not significant for AI \textit{w/ explanations} ($p=0.76$)}) . No main effect was found. As illustrated in Figure~\ref{fig:overlap}, these interaction effects show that for participants interacted with AI Assistant \textbf{\textit{w/o explanations}}, they accepted \textbf{more AI's suggestions} and \textbf{more false positive AI suggestions} in the \textbf{\textit{sync}} than \textbf{\textit{async}} condition. This difference, however, was not found for participants interacting with AI Assistant \textbf{\textit{w/ explanations}}. \rv{Figure~\ref{fig:overlap} also suggests participants interacting with AI \textit{w/ explanations}}  generally accepted more AI suggestions, in both \textit{sync} and \textit{async}, than the baseline condition (\textit{w/o, async}).

What's more, Table~\ref{tab:overlap} also suggests that AI's FPs and FNs had different effects on participants' analysis. In all conditions, the percentages of FPs leading to confirmed UX problems were low; this means that in most cases participants were able to recognize and reject AI's false suggestions. The percentages of FNs leading to confirmed UX problems were also low; this suggests that participants tended to neglect a UX problem if they were not reminded by the AI. These patterns reveal that in the context of identifying UX problems, FNs seem to be more harmful than FPs. The implication is that \rv{when designing a human-AI collaboration tool for UX evaluators, \textbf{it would be desirable to minimize FNs (i.e., the AI does not highlight the problem but it was actually there) over FPs (i.e., the AI highlights a problem but it was actually not there)} if trade-off between the two have to be made, for example by setting the classification threshold to have high recall.}

%\st{it might be desirable to \textbf{optimize \ FNs over FPs, if trade-offs between the two have to be made, when designing a human-AI collaboration tool for UX evaluators}}. 

Interestingly, after removing 3 outlier data points, we found that for Pct. AI FN Found, there is a marginally significant positive main effect of explanations ($\beta=0.16, SE=0.08,F(1,15)=4.18, p=0.06$; M(\textit{w explanations})=13.3\%; M(\textit{w/o explanations})=8.3\%). This implies that participants interacted with AI Assistant \textbf{\textit{w/ explanations}} were able to \textbf{find UX problems \rv{more effectively}} even if AI failed to remind them.

\subsubsection{Behavioral Patterns (RQ1, RQ2)}
We examined behavioral metrics that reflected how participants interacted with the tool, including the time spent analyzing a video, number of times they jumped forward in the video, rewound, paused, and the average time of each pause. Since the length of video might impact these behaviors, we normalized all metrics, except the average time per pause, by the length of the video (3.57 minutes for the website video and 11.37 minutes for the coffee machine video). That is, a normalized metric represents a participant' frequency of engagement in a corresponding behavior per every minute of a video. 
% \vera{I made a decision to normalize the behavioral measure. Since we normalized the count of problem, it would be consistent. Also, double check the experiment design description about giving participants twice the time, clearly, from the first column, they had more than twice the time of video} 
Table~\ref{tab:interaction} shows descriptive statistics of these metrics.

\begin{table}
  \caption{The behavioral metrics reflecting participants' interactions with AI Assistant (with means and standard deviation in parenthesis), including the total time spent, number of jump forwards, rewinds, and pauses, all of which are normalized by the length of the video (in minutes); and average duration of a pause (in seconds). }\label{tab:interaction}
 {\footnotesize
  \begin{tabular}{ccp{1.8cm}p{2.2cm}p{1.8cm}p{1.8cm}p{1.8cm}}
  %\begin{tabular}{Cc^c^c^c^c^c}
  
    \toprule
    \rowstyle{\bfseries}
       & & \textbf{Total time (normalized)} & \textbf{N Jump Forward (normalized)} & \textbf{N Rewind (normalized)} & \textbf{N Pause (normalized)} & \textbf{M Pause time}\\
    \midrule
    &\textit{async}&2.68 &0.57 &1.67 &2.13 &21.7 \\
    \textit{w/} &&(0.24) &(0.26) &(0.44) &(0.38) &(2.85) \\
    & \textit{sync}&2.21 &0.16 &1.25 &1.61 &18.1 \\
   & & (0.23) &(0.09) &(0.43) &(0.30) &(2.34) \\

    \midrule
    & \textit{async}&2.15 &0.18& 1.02 &1.78 &13.7 \\
    \textit{w/o}& &(0.32)&(0.09) &(0.34) &(0.40)  &(2.49) \\
    & \textit{sync} &2.83&0.30&1.51 &1.70 &17.6 \\
    & &(0.41)&(0.09) &(0.43) &(0.32) &(2.71) \\

  \bottomrule
\end{tabular}
}
\end{table}

\begin{figure*}[ht]
    \centering
\begin{subfigure}[b]{.47\textwidth}
  \centering
  \includegraphics[width=\linewidth]{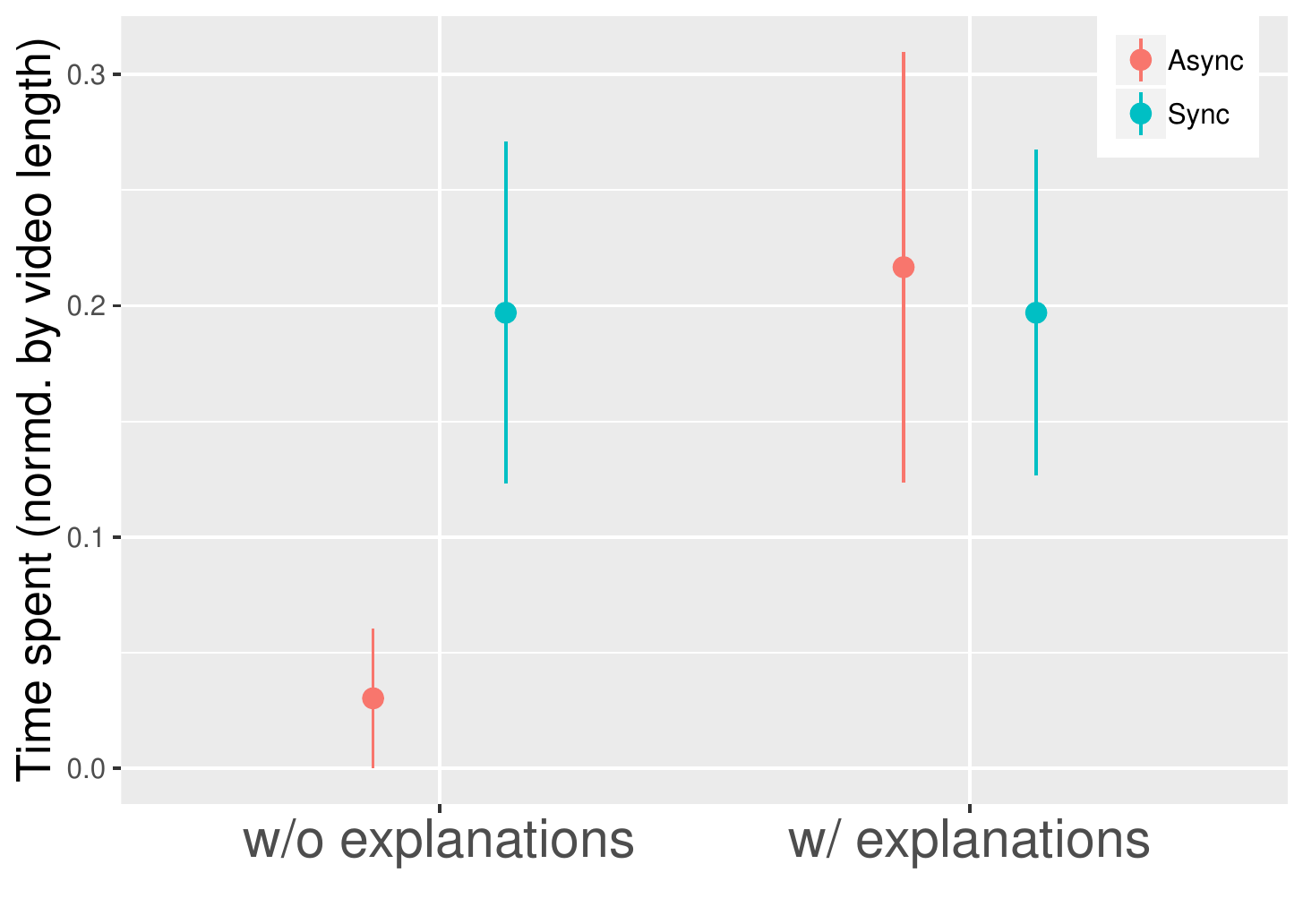}
 % \caption{Customer profile presented in all conditions for annotation}
 % \label{fig:interface_1}
 % \label{fig:sub1}
\end{subfigure} %
\begin{subfigure}[b]{.47\textwidth}
  \centering
  \includegraphics[width=\linewidth]{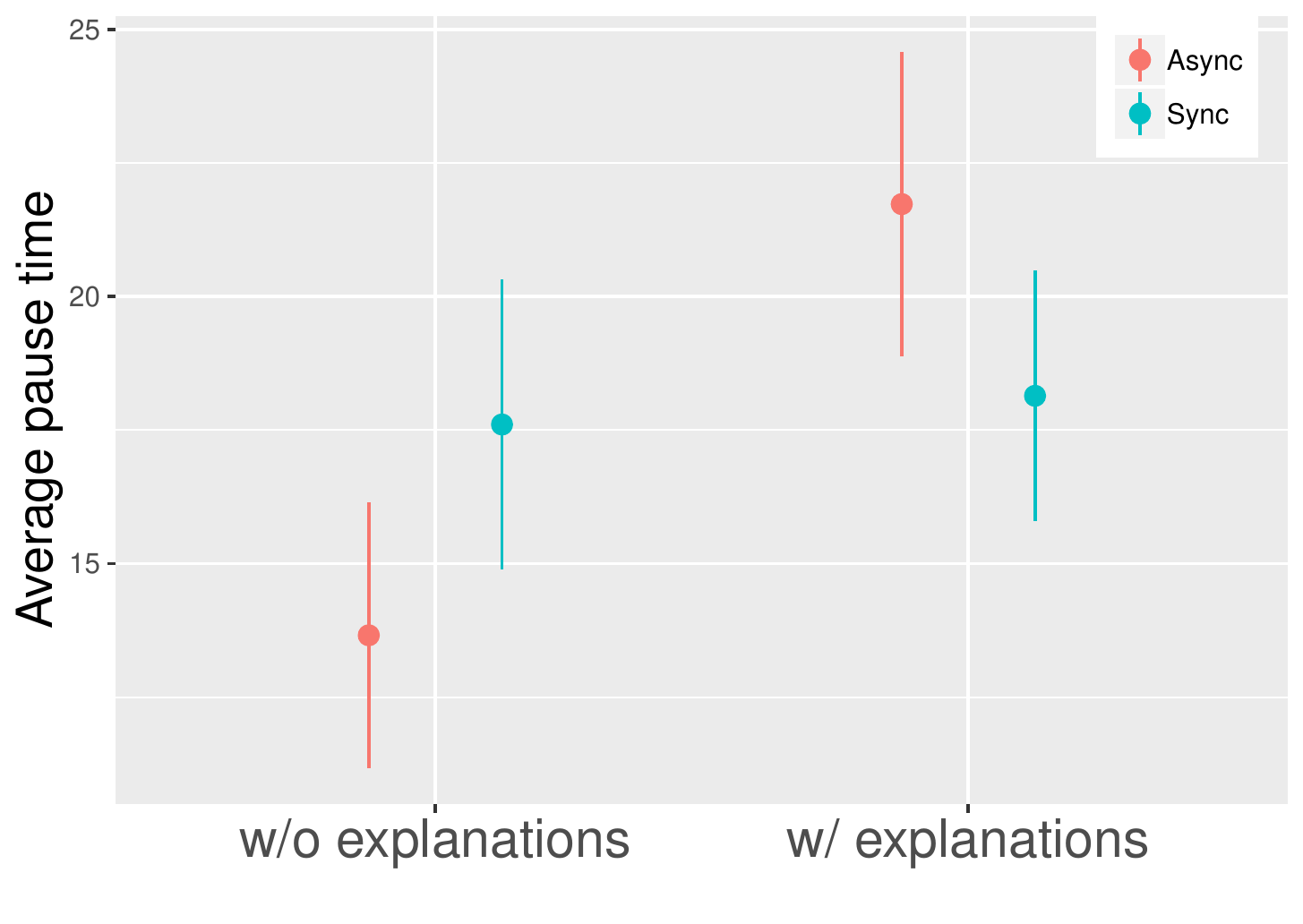}
  %\caption{explanations and questions presented in the XAL condition}
  % The final caption as appearing in the pdf needs reformating. Instead of having two different captions, you can have a single caption where you can manually specify (a) and (b). 
\end{subfigure}
    % \caption{Number of UX problems identified, precision, recall, and found UX problems overlapping with AI suggestions by participants across conditions. All error bars represent +/- one standard error.}
        \caption{Two-way interaction between explanations and synchronization on the time spent on analysis (normalized by video length, and average pause time, as shown by means and standard divisions across conditions. All error bars represent +/- one standard error. }
    \label{fig:behavior}
\end{figure*}

We performed a separate mixed-effects regression on each of these behavioral metrics. 
We found a significant interaction effect between the presence of explanations and synchronization on the normalized total time spent ($\beta=-0.99, SE=0.40,F(2,18)=6.03, p=0.02$\rv{; post-hoc analysis found the contrast between \textit{async} and \textit{sync} significant for AI \textit{w/o explanations} ($p=0.01$), but not significant for AI \textit{w/ explanations} ($p=0.48$)})), and a marginally significant effect on the average pause time (after removing an outlier) ($\beta=-7.12, SE=4.05,F(2,17)=3.09,p<0.10$\rv{; post-hoc analysis found the contrast between \textit{async} and \textit{sync} marginally significant for AI \textit{w/o explanations} ($p=0.10$), but not significant for AI \textit{w/ explanations} ($p=0.38$)})). 
As shown in Figure~\ref{fig:behavior}, these interaction effects are consistent with the trend of the identified UX problems, Recall, and percentage of AI suggestions confirmed described in the last subsections:
When interacting with AI Assistant \textbf{\textit{w/o explanations}}, participants \textbf{spent more time and paused longer} in the \textit{sync} than the \textit{async} condition. Such a difference did not appear for participants interacting with AI assistant \textit{w/ explanations}. Although not statistically significant, the same trend of interaction effect was observed for the number of forward jumps and rewinds as shown in Table~\ref{tab:interaction}. 

Taken altogether, these behavioral patterns suggest that when interacting with AI Assistant \textbf{\textit{w/o explanations}}, participants appeared to be \textbf{more engaged} and examined the usability test videos \textbf{more actively} in the \textit{sync} than the \textit{async} condition. 
 As a result, they found more UX problems in the \textit{sync} than the \textit{async} condition. For those interacting with AI Assistant \textbf{\textit{w/ explanations}}, their engagement and performance did not show significant difference between the \textit{sync} and \textit{async} conditions. \rv{Figure~\ref{fig:behavior} suggests participants interacting with AI \textit{w/ explanations} were generally \rv{more engaged} (spent more time and paused longer) than the baseline condition \textit{(w/o, async)}}.

\subsubsection{Subjective Perceptions (RQ1-3)}

We analyzed the measurements of the subjective perceptions of the tool from questionnaire responses. After analyzing videos with each version of AI Assistant (\textit{sync} or \textit{async}), participants completed a questionnaire rating their \textit{Satisfaction}, \textit{Understanding} and \textit{Trust} in AI Assistant, as described in Section~\ref{measure}. Table~\ref{tab:survey} shows the descriptive statistics of these questionnaire responses.
\begin{table}
  \caption{The subjective perceptions (e.g., satisfaction, understanding, and trust) of AI Assistant reported in the post-test questionnaire, with mean values and standard errors in parenthesis}\label{tab:survey}
 {\footnotesize
  %\begin{tabular}{p{1.5cm}p{1.7cm}p{1.5cm}p{1.5cm}p{1.5cm}p{1.5cm}p{1.5cm}p{1.6cm}}
  \begin{tabular}{Cc^c^c^c^c}
  
    \toprule
    \rowstyle{\bfseries}
        & &Satisfaction&Understanding&Trust\\
    \midrule
    &\textit{async} &5.53 &5.90 &4.80  \\
    \textit{w/}& &(0.23) &(0.22) &(0.40)  \\
    &\textit{sync} &5.48 &5.77 &4.94  \\
   & &(0.36) &(0.26) &(0.38)  \\

    \midrule
    &\textit{async} &4.88&4.64 &4.55  \\
    \textit{w/o}&&(0.42) &(0.47) &(0.40)   \\
    &\textit{sync} &5.39 &4.82 &4.88 \\
    & &(0.25) &(0.44) &(0.29) \\

  \bottomrule
\end{tabular}
}
\end{table}

Separate mixed-effects regression models on the three subjective perceptions found a main effect of explanations on perceived Understanding ($\beta=0.92, SE=0.44,F(1,15)=4.35, p=0.05$), suggesting that participants found the explanations helpful for them to better understand how AI Assistant worked. While not statistically significant, the measure of satisfaction exhibits a consistent trend with the results of task performance and behavioral patterns as indicated in previous subsections ($\beta=-0.55, SE=0.47,F(2,18)=1.36, p=0.26$). The main effect of explanations is also trending significant ($\beta=0.68, SE=0.52,F(2,15)=1.73, p=0.20$). That is, for participants interacting with AI Assistant \textbf{\textit{w/o explanations}}, they felt \textbf{\rv{somewhat }more satisfied} in the \textit{sync} than the \textit{async} condition. For participants interacting with AI Assistant \textbf{\textit{w/ explanations}}, they felt they \textbf{understood the AI better}, and \textbf{were somewhat more satisfied} with the tool, in both the \textit{sync} and \textit{async} conditions, than the baseline condition \textit{(w/o, async)}. 

% \vera{I am not sure whether we should tone down about the conclusion on satisfaction above, since they are not statistically sig. say something like ``trending more satisfied'' or ``somewhat more satisfied''?}

It is worth noting that as controlled variables in regression models, we found a main effect of general Trust in AI in increasing Satisfaction ($\beta=0.65, SE=0.36,F(1,15)=0.09, p=0.09$), Understanding ($\beta=0.99, SE=0.36,F(1,15)=7.93, p=0.01$) and Trust ($\beta=0.97, SE=0.39,F(1,15)=6.09, p=0.03$). This indicates that participants with a generally positive attitude towards AI also had more positive perceptions of AI Assistant.

\rv{In summary, we found that synchronization had a significant effect for participants interacting with AI Assistant \textit{w/o explanations}:  \textit{sync} AI led to  higher engagement (more time spent and longer pause), more acceptance of AI suggestions, better performance of finding UX problems, and somewhat higher satisfaction with the AI Assistant. Such difference from synchronization was not found for the group interacting AI Assistant \textit{w/ explanations}, who were also found to have a better perceived understanding of AI Assistant, higher ability to detect usability problems from the video even if the AI Assistant failed to remind them. \st{, and incorporate explanation content into their problem descriptions}. The latter group also exhibited a tendency of higher engagement and more acceptance of AI suggestions compared to the baseline condition \textit{(w/o, async)}}.

% \subsubsection{Key Takeaways}.

% without explanations, sync is better.
% find quotes to find out why they thought sync was good.

% 5.2.1., understanding interaction effect. 
% ---without explanations, find quotes to support sync was better than async

% explanations is good... regardless of sync
% ---

%% file: texfiles/6-qualitativeresults.tex
\subsection{Qualitative Results}
\label{qualitative_results}

\rv{Based on the quantitative results above, we first discuss insights from qualitative data to further unpack the benefits of AI explanations and synchronization, and how they might have worked together to create the interactive effect on participants' engagement and performance. }We will also discuss themes on the general perception of AI Assistant \rv{to support UX evaluators to perform usability test video analysis} (Section~\ref{sec:sub_generalObservationAI}) and \rv{suggest} ways to improve the design (Section~\ref{sec:sub_waystoimproveAI}). We will highlight the themes identified from the thematic analysis in \textbf{bold} and support them with selected quotes. \mr{To help readers contextualize the quotes, we annotated them with their corresponding Explanation conditions \textit{(w vs. w/o)} since Explanation was a between-subjects factor and each participant experienced only one of the two conditions. In contrast, since Synchronization was a within-subject factor and all participants experienced two conditions \textit{(sync vs. async)}, we only annotated ones that were clear which synchronization condition they referred to.}

%First, we present qualitative data to better understand the quantitative results about synchronization (Section~\ref{sec:sub_synchronization}) and explanations (Section~\ref{sec:sub_explanations}). Then, we highlight the general perceptions and usage of AI (Section~\ref{sec:sub_generalObservationAI}) and ways to improve AI for human-AI collaboration (Section~\ref{sec:sub_waystoimproveAI}).
\subsubsection{\rv{Benefits of} Explanations (RQ1)}
\label{sec:sub_explanations}
%\st{A key finding of our quantitative analyses was that when AI explanations was presented, participants performed better than the baseline condition (\textit{async, w/}) regardless of synchronization. They also found more false negative UX problems even when the AI did not suggest and reported significantly better understanding of AI Assistant.Our qualitative analysis revealed three main advantages of having access to AI's explanations.}

\rv{First, we found further support consistent with the quantitative results that} 
%\st{First, consistent with quantitative results on self-reported understanding  a direct effect of} 
having access to AI's explanations led to a perceived \textbf{better understanding} of how the AI works.
\begin{quote}
    ``\textit{I felt that I could rely on it more than half of the time. It detected what the participant said and did. By combing these two factors, I felt it often generated reasonable suggestions.-P18\rv{, w/}}'' 
\end{quote}

%\st{Second, the improved understanding could have led to \textbf{higher trust} in the AI. This is illustrated by the lower frequency of mentioning a lack of trust among participants who used the AI \textit{with} explanations (3 out of 12) than those who used the AI \textit{without} explanations (9 out of 12):}

\rv{Despite a lack of significant survey results on user trust, we found some evidence that, by having explanations, the \textbf{improved user understanding can potentially improve trust} in AI Assistant. This is illustrated by the lower frequency of mentioning a lack of trust among participants who used the AI \textit{with} explanations (3 out of 12) than those who used the AI \textit{without} explanations (9 out of 12), for example:}

%\st{I don\'t feel I have faith in the AI. I would definitely want to get more trials to see if they really work before deciding whether to use the AI\'s inferences or not.-P24, w/o}

\begin{quote}
    
    ``\textit{Maybe after I work with the AI for a while and really feel it helps me identify problems, I will build trust with it. But for now, since I feel I don't know enough about the AI and I don't know how it works, I am not sure I can 100\% trust it.-P23, w/o}'' 
\end{quote}

%\st{Third participants found the content of explanations to have \textbf{utility for UX evaluation tasks}. Even if they didn't completely trust the AI's capability, they found reading the explanations helped them make more informed judgments of UX problems:}

\rv{Second, participants felt the explanations had \textbf{utility for UX evaluation tasks}, which could help them make more informed judgments of UX problems: }

\begin{quote}
    ``\textit{I trust myself more than the AI. For example, while the AI thought the user encountered an issue when searching for the instruction picture, I think it was just a typical searching process and was not necessarily an issue. But I think it is valuable to review what the AI says and why and then make my own decision.-P23\rv{, w/o}}'' %-P31 wo
\end{quote}

%\st{More specifically, participants used the explanations content to remind themselves of the indicators of potential UX problems that they might have otherwise missed.}
\rv{Consistent with the observations that participants exhibited better ability to detect UX problems and higher engagement when explanations were presented, many commented on \textbf{attending to the explanation contents}:}
%help them better allocate their attentions when reviewing the video, or
\begin{quote}
    ``\textit{I like the tool was able to catch some very detailed stuff because I was not able to divide my attention to multiple parts of the whole video like both the participant face video and the screen capture.-P17\rv{, w/}}'' %-P25 w
    
    ``\textit{I was bad at detecting the uncertainty in the user's tone and hoped that the tool was able to pick that up.-P20\rv{, w/o}}'' %-P30 wo
    \end{quote}

This utilization of explanations content is also demonstrated by participants' tendency to incorporate the content in AI's explanations in their submitted UX problem descriptions, as discussed in the quantitative analysis shown in the last column of Table~\ref{tab:overlap}.

%\ming{updated the table reference. make sure this is correct.}
% In contrast, when using the AI without explanations, participants expressed their doubts about the AI. 
% They wondered more often about how the AI worked and needed to work with the AI for more trials to build trust in it. 
%\vera{perhaps tone down the paragraph below because it is a very strong argument to say without explanations no trust if you do not provide detail on how you did the comparison. maybe ``there is evidence that without explanations, more participants expressed hesitation to trust the AI''}
%Participants had different attitudes toward the AI Assistant \textit{w/} and \textit{w/o} explanations. 

%\vera{to-do: come back to work on this paragraph after deciding on framing}
%The above qualitative findings  explanations could increase AI agency in the human-AI collaborative task, not just by improving user's understanding and trust, but also actual agency with additional task support by providing useful rationales for identifying UX problems in the test videos. These benefits may explain participants' improved analysis, especially their ability to identify more false-negative problems, even when the AI did not explicitly suggest.  

\rv{The above qualitative findings highlight the benefits of providing AI explanations in assisting usability video analysis. Not only did they help users better understand the AI, but also provided additional task support to help identify usability problems. These benefits may explain participants' improved analysis, especially their ability to identify more false-negative problems, even when the AI did not explicitly suggest. Interestingly, we did not find a compromise of human agency by providing explanations, as suggested by the notion of human-AI agency divide~\cite{lai2019human,Lai2020Why}, but instead found improved human engagement in analyzing the video as compared to the baseline condition. }

%However, with explanations, participants tended to have a better understanding of how the AI made suggestions.

%Third, although participants might not take AI's suggestions, they felt it was valuable to consider AI's explanationss before making informed decisions.

\subsubsection{\rv{Benefits of }Synchronization (RQ2)}

\label{sec:subsubsynchronization}
% \todo{the story will be much more complete if you can show the count of complaints about async AI being biasing, distracting, overwhelming is higher in without explanations conditions }
\label{sec:sub_synchronization}
A key finding of our quantitative analyses was that when explanations were not present, participants better engaged with the videos and performed significantly better when the AI suggestions were provided \textit{synchronously} than {asynchronously}. Our qualitative data revealed four main reasons how the \textit{sync} AI helped participants.

First, participants felt that the \textit{sync} AI allowed them to better perform \textbf{independent analysis}, and not be biased by the amount and places of UX problems in a video, as illustrated by the following quote from a participant interacting with AI Assistant \title{w/o explanations}.
%Participants did not want to be influenced before making their own judgment on the amount and places of, and seeing the AI-suggested problems could potentially bias their own judgment.
\begin{quote}
 %   ``\textit{ It's crazy to know all the (\textit{async} AI's) suggestions before I analyze the video. I want to have my own first impressions and judgments because I think that's really important.-P13, w}'' % w
    
    ``\textit{I don't prefer it (the async AI) showing me many indications of problems before I make my own judgment. I don't want to be biased.-P24, (w/o, async)}''%P32 wo

\end{quote}

Second, compared to the \textit{async} AI, which showed the AI's suggestions all at once since the beginning of their analysis, the \textit{sync} AI was \textbf{less overwhelming}. 
\begin{quote}
    ``\textit{Async [AI] is a bit overwhelming with the problems revealed all at once. It could bias me before I am able to look into the video and have my judgments.-P20, (w/, async)}'' %P30 wo
\end{quote}

Relatedly, participants found the \textit{sync} AI allowed them to \textbf{better focus} on their current analysis without being distracted by AI's suggestions for the forthcoming portion of the video.

\begin{quote}
    ``\textit{[With] the sync [AI] I can pay more attention to what's going on right now. But the async [AI] also tells you the next usability problem that is going to happen. In that case, I found myself tend to focus more on what's going to happen next instead of finding the issue that is happening now. So it distracted me from my current analysis.-P19, (w/o, sync)}''
\end{quote}

Last but not least, the \textit{sync} AI created a stronger sense of \textbf{social presence},  with the perception of analyzing the video with another ``evaluator'' simultaneously. 
\begin{quote}
    ``\textit{It feels like another person is doing this [analysis] with me.-P1, (w/, sync)}''% w
\end{quote}

Although the above results suggest \textit{the \textit{async} AI allowed lower human agency}, participants also pointed out two potential benefits.
First, the \textit{async} AI allowed them to get an \textbf{overview} of the distribution of the potential UX problems in the video. 
Second, the \textit{async} AI resulted in \textbf{less frequent visual updates} on the interface than the \textit{sync} AI, therefore was less demanding on their attention and cognitive resources. 

\begin{quote}
    ``\textit{I really like the AI highlighted all the problems at the beginning so that way I could pay attention to certain parts of the video. In that way, I could get things done much faster.-P15, (w/o, async)}''% wo
    
    % ``\textit{The problems kept popping up [in the sync AI], which distracted me from watching the video.-P18}'' %P26 w
\end{quote}
%Fourth, the \textit{sync} AI allowed them to focus on their current analysis better because it did not show AI's judgments for the upcoming portion of the video. In contrast, they felt being distracted from their current analysis when seeing the \textit{async} AI's judgments for the forthcoming portion of the video.

\rv{\textit{The observations in the last two sub-sections can help interpret the interactive effect between synchronization and explanations}. For the AI Agent \textit{w/o explanations}, participants repeatedly expressed a level of mistrust in its capability. The synchronous AI, which progressively reveals its suggestions as the evaluator examines the video, validating their analysis rather than dictating the job, might have been a better match than the asynchronous AI to support a desired level of human agency. As a result, UX evaluators had better engagement in analyzing the video and better analysis outcomes.

This mismatch between the asynchronous AI and UX evaluators' desired human agency might have been mitigated with the explanations provided by the AI Assistant, which could have been perceived to be more capable and trustworthy. Perhaps more importantly, the content of explanations provided additional information and utility for evaluators to analyze the video, which could have helped them better identify usability problems and stay more engaged regardless of the synchronization, as the quantitative results showed.}

\subsubsection{ \rv{Perceived Values }of AI Assistant \rv{for Usability Video Analysis} (RQ3)}

\label{sec:sub_generalObservationAI}
%\st{Our qualitative results also revealed participants' general perception and usage of AI for usability video analysis. Participants felt that the AI was helpful in four ways.} 
\rv{To answer RQ3, we summarize the key values that participants saw from the WoZ AI Assistant to inform effective forms of human-AI collaboration for usability video analysis.}
First, \rv{participants appreciated} AI to provide a \textbf{second opinion to verify} their analysis, which UX practitioners often \rv{desire but would be expensive to obtain from another human evaluator}  since most times they perform usability analysis alone~\cite{fan2020Survey,folstad2012analysis}. 
\begin{quote}
    ``\textit{I can compare what I think is a problem to what the AI thinks is a problem. It gave me a second opinion without calling for a second designer for help.-P11, w/o}'' % wo
    
    ``\textit{Oftentimes one researcher has to cover several projects, so it's certainly nice to have a second opinion and also it can correct me and identify something I miss.-P23, w/o}'' %-P31 wo
\end{quote}

Interestingly, some felt that the AI Assistant was like a ``junior colleague'' \rv{or \textit{``a new intern who needs to be monitored and costs time and energy''} (P17). Given the performance limitation of AI and subjectivity of usability video analysis, this could be a proper mental model to design for effective human-AI collaboration in this context. We may argue that \textit{sync} AI,  without dominating the analysis, is a better fit for such a mental model. It would also be interesting to explore retrospective AI that reveals its suggestions after the evaluators' judgment. }

%\st{, who is not yet as competent, and their opinions should be taken with caution. }

%st{ [The AI is] just like a new intern who needs to be monitored and costs time and energy.-P17}

%\vera{not sure point 2 and point 3 differ, both regarding efficiency? consider rephrasing point 2}
Second, participants desired for the AI to \textbf{accelerate their analysis} by suggesting where to focus on in the video, especially when the video is long or when UX evaluators work under time constraints.
\begin{quote}
    ``\textit{The AI might speed up the whole evaluation, especially with time constraints. I can use it, to some degree, to accelerate my analysis by finding which parts I should go a little deeper.-P20, w/o}'' %-P30 wo
\end{quote}

\rv{However, our quantitative analysis revealed the risk of being misled by AI's false negatives by following AI's suggestions to navigate the video. Future work should balance the efficiency aspect and users' engagement with analyzing video content. }

%\st{One way of accelerating the analysis is to utilize AI's suggestions to better navigate the video, such as skipping parts of it to save time.}

  %  ``\textit{It's helpful for me to see what parts of the video I should pay attention to.-P15}'' % wo
    
%\st{I would be lazy and won't look through the entire video with the help of this AI, which could be just my preference, but I like it.-P11}

Third, the AI could \textbf{complement} participants' analysis by identifying issues that they might otherwise miss. As discussed \rv{in Sec.~\ref{sec:sub_explanations}}, many participants found the explanations particularly helpful for directing their attention to details that they did not recognize.  
\begin{quote}
    
   ``\textit{It gave me suggestions before I found issues and also reminds me of the places that I did not recognize the issue.-P12, w/}'' % w
\end{quote}

Finally,  \rv{some participants used AI's suggestions, such as the timeline feature, as } \textbf{anchor for further analysis} or to know where to replay the video.
%assisted participants with locating places when they replayed the video. 
\begin{quote}
    ``\textit{When I went back to watch the video for the second time, I knew where to pay attention to with the visualization of the problems.-P15, w/o}'' % wo
    
    ``\textit{When I double-checked my analysis by playing the video back and forth, it helped me locate the portions where usability problems might exist.-P13, w/}'' % w
\end{quote}

% While participants felt AI was helpful in the above five ways, it does not mean that they would always agree with the AI.  

\subsubsection{Ways to Improve the Collaboration between UX Evaluators and AI (RQ1-3)}
\label{sec:sub_waystoimproveAI}
\label{ways-to-improve-ai}
Participants also expressed ways in which the AI could be improved to better assist them with their analysis. %in terms of \textit{explanations} and \textit{synchronization}.

\textbf{Explanations.}  
The current AI Assistant describes five behavioral features in the visual and audio of a usability test video to explain the AI's suggestions. 
However, participants had different individual preferences for the importance of these features, and desired for ways to \textbf{personalize the rank and types} of features provided.

%and thus suggested to rank the features in a different order to better match their perceived importance or relevance of the features. Moreover, they also wanted to be able to make some features optional or add additional features. 
\begin{quote}
    ``\textit{I would rank the five features in this way: what the user said, what the user did, sentiment, tone, and speech rate because i felt tone and speech rate were not as important as the other three.-P13, w/}'' % w
    
    ``\textit{ I would suggest making speech rate optional. It's not critical for decision making.-P24, w/o}'' %-P32 wo
    
%    ``\textit{I don't need the AI to provide me things like emotion because I can feel it by myself.-P16}'' % w
\end{quote}

Participants also had \textbf{different preferences for when} the explanations should be shown. 
While some suggested the explanations to be shown only when they request them, others prefer seeing the explanations all the time instead of just the times when the AI detects problems.
\begin{quote}
    ``\textit{I hope the AI can show the information when I need it instead of providing information to me all the time.-P13, w/}''

    ``\textit{Instead of presenting this dialogue of information only during the time when the AI thinks there is a usability problem, maybe just show it through the video, which would be easier for me to analyze the problems.-P16, w/}''
\end{quote}

\textbf{Synchronization.} 
% \todo{this sub-section is a little tricky as some points are not really about timing, but improving AI or AI suggestion...let's discuss tmr}
Participants suggested three alternative designs to consider the timing to reveal AI suggestions.
The first approach is to allow UX evaluators to analyze the video by themselves first, and then the AI reveals its suggestions to help verify judgments in the \textbf{second pass}. 
\begin{quote}
%    ``\textit{I would like to review the video by myself and take my own notes first, and then I could compare my notes with the AI's suggestions.-P16}'' % w
    
    ``\textit{I prefer analyzing the video by myself first and then take a second look at it, and that is when the AI can show me the result because I do want to get a different perspective on what's going on in the video.-P24, w/o}'' %-P32 wo
\end{quote}

Second, 
%The revealing of the AI-suggested problems was agnostic to whether the UX evaluator completed her analysis for that particular time or not.
some participants suggested \rv{retrospective AI, by revealing} its suggestions \textbf{immediately after the UX evaluator indicates their judgment}, for example, by showing agreement or flagging missed problems. 
Such designs would further avoid biasing the UX evaluator's judgments. 
\begin{quote}
    ``\textit{I think it would be more helpful if the AI can analyze with me at the exact same time. For example, it shows me its inference at the exact same time when I find an issue.-P12, w/}'' % w
\end{quote}

%\vera{how is this different from async?}
%\todo{I still don't quite understand this approach}
% The third approach is to let the AI analyze the video first and then UX evaluators would check the result before using it. 
% This is an alternative way of asynchronously revealing AI-suggested problems. 
% \begin{quote}
%     ``\textit{After the AI automatically generates the result, i will go through it myself to make sure I am OK with the result.-P23}'' %-P31 wo
% \end{quote}

The third approach is to let the AI analyze a video asynchronously but \textbf{flag suggestions that the AI is uncertain about}. In this way, UX evaluators could save time on the problems that AI is confident about and focus their attentions on the ones that AI would likely make mistakes.

\textbf{Additional Features and Functions.} 
\begin{comment}
\todo{is this one still about explanations? if so we need a more explicit summary sentence (think of it as a code) if not we can also remove this part}
\end{comment}
Participants also suggested additional features and functions. 
First, participants expressed the desire to know \textbf{richer information about AI}, such as the accuracy of the AI and the principles that the AI uses.
Moreover, they wanted to see more information about UX problems, such as the type, frequency, and severity of the suggested problems, and perhaps even recommendations for fixing the problems.    
\begin{quote}
    ``\textit{I don't know the authenticity of the AI, such as its accuracy...It's nice to have many types of principles, not only Nielsen's, for the AI to apply and have an option to select a set of principles for the AI, which might generate different results.-P11, w/o}'' % wo

    ``\textit{Instead of presenting this dialogue of information only during the time when the AI thinks there is a usability problem, maybe just show it through the video, which would be easier for me to analyze the problems.-P16, w/}'' % w
\end{quote}

Second, they hoped that the AI could adapt its inferences by \textbf{learning from human guidance}. 
For example, UX evaluators could analyze a few usability test sessions first, and then the AI learns their analysis and automatically analyzes the remaining sessions.
A different approach is to let UX evaluators review and edit the AI's suggestions for the AI to learn from. These active learning approaches could allow the AI to be better customized to detect UX problems in specific technology domains.
% \begin{quote}
%     ``\textit{I hope I can have the option to edit AI's suggestions and train it so that it can be closer to my judgments.-P11}'' %wo
% \end{quote}
Furthermore, they suggested another way to improve AI is to allow for choosing different UX design principles, other than Nielsen's heuristics~\cite{nielsen2005ten} as used in this work, for the AI to consider. 

Last, participants also hoped the AI could generate \textbf{a report of UX problems}. For example, the AI could generate the description for each problem, which UX evaluators could either directly use or edit further if needed when they create their UX analysis report. 
\begin{quote}
    ``\textit{The AI may provide some problem descriptions, so we just need to check the box to select or edit instead of having to type the descriptions all by ourselves. It might make the analysis much faster.-P12, w/}'' % w
\end{quote}

%% file: texfiles/7-discussion.tex
\section{Discussion}
\label{sec:discussion}
To summarize, we started from the baseline design of AI Assistant, which was the asynchronous AI without explanations \textit{(w/o, async)}, and extended it into three additional designs of AI by changing two factors--explanations and synchronization. We found that: 
\begin{itemize}
\item \textbf{Explanations}: \rv{When the AI Assistant provided explanations, participants performed equally well regardless of synchronization, and exhibited higher engagement and more acceptance of AI suggestions compared to the baseline AI \textit{(w/o, async)}. Explanations also improved participants' self-reported understanding of the AI, ability to detect usability problems even if the AI Assistant failed to remind them \st{, and allowed incorporation of explanation content into their usability problem reports}}. 

\item \textbf{Synchronization}: When without explanations, \textit{Synchronous} AI improved UX evaluators' performance and engagement with AI's suggestions more than the baseline AI \textit{(w/o, async)}.
\end{itemize}

Next, we first elaborate on these findings and their implications for designing AI explanations and synchronization, and then discuss high-level takeaways regarding human-AI collaboration and AI for UX work.

\subsection{Design Implications: Explanations and Synchronization}
While our quantitative results suggest both explanations and synchronization could benefit human-AI collaboration in the context of UX evaluation, our qualitative data further reveal the underlying reasons and suggest design implications

\textbf{Explanations.}
Consistent with prior work~\cite{cheng2019explaining,kocielnik2019will,buccinca2020proxy}, we found that providing explanations could robustly improve users' perceived understanding of AI. More importantly, participants appreciated the additional support enabled by explanations, to better direct their attention to the parts of the video likely associated with UX problems, and to remind them of the indicators of UX problems that they might have otherwise missed. Some participants even wished to have constant access to the explanation of the AI's input features, suggesting that they did not merely view these features as supporting AI's suggestions, but also utilized them in their own analysis of the videos.

%which demonstrated that by providing explanations, AI could provide additional agency and relieve humans from certain cognitive efforts to process information or reason about the decision. 
% \todo{is there any qual evidence that participants commented on the utility or preference between the heuristic part and feature part of the explanation? I think that would be a good design implication to discuss}

Our WoZ design of AI explanations provided two pieces of information to support UX evaluation: violations of design principles or heuristics~\cite{nielsen2005ten} (rule-based explanations), and behavioral features of the test subjects that are potential indicators of UX problems~\cite{fan2019concurrent} (feature-based explanations). Future work could explore technical approaches to generating such explanations in a robust way.

\st{Another potential utility of AI explanations, as revealed in the interview feedback and overlaps between AI's explanations and participants' problem descriptions in the last column of Table~\ref{tab:overlap}, is that participants incorporated the information in explanations into their problem descriptions. As UX practitioners often write and present UX analysis reports to the product team~\cite{fan2020Survey,folstad2012analysis,mcdonald2012exploring}, future work could explore ways to automate or assist with UX report generation using AI explanations. In such a collaborative UX report generation process, it is worth exploring whether and how to delineate the responsibilities between the human and the AI.}

Further, our qualitative data reveal varying individual preferences for the types and ranks of AI's features to be shown, and a desire to customize the explanations. 
%As Sec~\ref{sec:sub_explanation} shows, 
It is worth exploring ways to balance two perspectives of using explanations: as \textit{justification} of AI's decisions to better understand the AI and as \textit{additional information} to better support the UX evaluation task. Current XAI methods usually adopt the former perspective. A fixed set rules or features are shown as explanations based on how the AI actually arrives at its decisions. There is a growing criticism that this algorithm-centric view may miss great opportunities to support a primary motivation why people seek explanations~\cite{miller2019explanation}: a joint sense-making process where both the explainer and explainee supply information or constraints to build common ground. We believe knowledge-rich domains where experts and AI collaborate, such as the usability video analysis task we studied, provide interesting test grounds to explore novel interactive explanation techniques that allow experts to set preferences or constraints, even provide feedback or critique, for AI to improve its explanations so as to better support the experts' tasks.

\rv{Lastly, we should point out that recent works warned that explanations could lead to unwarranted trust~\cite{poursabzi2018manipulating,Ghai2020XAL,bansal2021does,zhang2020effect} and over-reliance on AI, if a user simply associates explanability with AI capability without engaging analytically with the model behaviors. While our results did not suggest this is a salient problem, future work introducing real, imperfect AI technologies should be mindful about such a risk. For example, recent studies suggested adaptive explanations---only providing explanations for suggestions that the AI is confident about~\cite{bansal2019updates,zhang2020effect}.}

\textbf{Synchronization.} In current AI-mediated decision support systems, it is common to adopt an asynchronous user mental model, in which the AI finishes the analysis first and presents its analysis to domain experts (e.g., ~\cite{fan2020vista}). Our study, however, reveals valuable benefits of providing synchronous AI support, unfolding in real time as the user performs the task. With the synchronous AI, our participants particularly welcomed their ability to retain high human agency, not get biased or distracted by AI's suggestions beyond the portion of the video that they have analyzed, and get a second opinion to verify or complement their judgment, which is often lacking in UX practice~\cite{fan2020Survey,folstad2012analysis} but important for overcoming the ``evaluator effect''~\cite{hertzum2001evaluator}.

Furthermore, the synchronous AI also helped to create a stronger ``social presence,'' the feeling of working together with a colleague, than the asynchronous AI. In CSCW literature, studies repeatedly found that workers exhibit higher engagement, creativity and social interactions in synchronous than asynchronous collaboration context\mr{s}~\cite{shirani1999task,birnholtz2012tracking}, while asynchronous collaboration may incite social loafing, with which an individual exerts less effort working with others. This risk could possibly extend to human-AI collaboration, as we observed less engagement and lower task performance in the baseline asynchronous condition than the synchronous conditions.

It is interesting that the comparative disadvantages of the asynchronous AI Assistant were mitigated if it provided explanations. We suspect that it may relate to users' mental model of the capability of AI Assistant. In the qualitative data, participants repeatedly mentioned that they did not place high confidence on the AI's capability to detect all UX problems in the videos, seeing it as a ``junior colleague,'' and preferred to rely on themselves and only utilized the AI's suggestions as verification or reminders. Thus, without dictating the analysis or distracting from their own judgment, the synchronous AI was a better fit for such a mental model than the asynchronous AI, which pushed all its judgements to them before they even started their own analysis. When the explanations panel was added (\textit{AI w/ explanations}), participants viewed the explanations as an additional utility feature to guide them better navigate and observe the actions in the videos, which may explain their equal engagement in both the synch and asynch AI conditions.

It is worth noting that the takeaway message is not that synchronous AI should always be preferred over asynchronous AI. Instead, this design decision should depend on the AI agency and the type of support that the AI can provide. It was the mismatch between the collaboration timing, where the AI dictates the analysis in the asynchronous condition, and the AI's limited range of support (i.e., without explanations) that led to the undesirable outcome. 
In knowledge-rich domains, such as UX evaluation, it is not uncommon that users place more faith in their domain expertise and judgments than the current AI and prefer to have a synchronous AI assistant. However, as users grow their trust in the AI, they might enjoy an asynchronous AI assistant that can provide an overview of the problems in the entire video as indicated in the last paragraph in Sec~\ref{sec:subsubsynchronization}. \mr{Alternatively, users might prefer the flexibility to toggle between synchronous and asynchronous collaboration with AI.} 
% in other tasks where the users are less of experts, or after they establish high confidence over AI's performance, users could prefer asynchronous or partially asynchronous AI, for example when using AI Assistant to analyze simple interfaces.

\subsection{\rv{Human Agency} and Timing of Human-AI Collaboration}

We would like to extend the point of designing for different Human-AI agency divide and matching the timing of Human-AI collaboration further. First, providing explanations is not the only way to increase the level of AI assistance, in terms of the quality and scope of support it can provide. Lai et al.'s~\cite{lai2019human,Lai2020Why} work suggested another type of information to increase AI agency--performance information to give users confidence and relieve them from the cognitive effort in judging the AI's recommendations. We found similar suggestions from participants to see richer information about the AI's performance and confidence as indicated in Sec.~\ref{sec:sub_waystoimproveAI}. 
%We also note that given the nature of usability video analysis tasks, where the types of interfaces and usability problems are diverse, it may not be advisable to provide a system-wide accuracy information. Instead, performance information at the individual decision level or groups of decision, such as confidence scores, success cases or limitations (for what systems or tasks the AI might fail), could be more effective in helping users deciding when to delegate the judgment to AI with less effort. 
The implication is that designers need to consider more nuanced designs for the timing factor of human-AI collaboration, for example, by allowing the AI to work asynchronously for high-confidence cases but synchronously for low-confidence cases.

There are several other frameworks on different levels of AI assistance
%There is a wide spectrum of machine/AI assistance 
for both broader automated systems ~\cite{parasuraman2000model,mackeprang2019discovering} and specific AI applications~\cite{young2007driving, lee2019human}. %built on a classic model of automation levels~\cite{parasuraman2000model}, 
Mackeprang et al. suggested a 10-level framework for different configurations of machine assistance, ranging from offering no assistance, to offering a set of suggestions, to automatically confirming with different ways for humans to intervene, to acting autonomously~\cite{mackeprang2019discovering}. Our current designs of AI Assistant were at the lower end of this automation spectrum, with the AI only offering suggestions, so the synchronous collaboration was more suitable than the asynchronous one. In the future, if the AI becomes more competent and could be confidently tasked with auto-detection and report-generation for usability video analysis, the timing factor should be reconsidered accordingly. For example, some participants suggested letting the AI perform the analysis asynchronously and then flag the decisions that it is uncertain about for human evaluators to double check. 

Last but not least, AI agency and human agency are not necessarily a dichotomous trade-off. \rv{On the one hand, we found some evidence that providing explanations improved participants' perceived AI capability, even trust. On the other hand, we did not find a compromise of human agency. To the contrary, explanations seemed to have ``slowed down'' the evaluators and improved their ability to detect usability problems even if the AI failed to suggest. Recent work by Shneiderman ~\cite{shneiderman2020human} highlights that human-AI agency is not necessarily a dichotomous trade-off. Instead reliable and safe AI design should strive for both high AI and high human agency to exercise effective control. Our results suggest that providing explanation could be a possible way to achieve this goal.}

\subsection{AI for UX Evaluation}
Our work provides empirical insights into UX professionals' attitudes, needs and preferences for AI technologies to support UX evaluation work. On the one hand, they were positive about the idea of having AI support to accelerate and complement their work, especially given that UX professionals often analyze a test session alone under time pressure~\cite{fan2020Survey,mcdonald2012exploring,folstad2012analysis}. They especially welcomed the idea of relieving them from tedious or mechanic part of their tasks, such as maintaining close attention to lengthy usability test videos or writing up usability reports. 

On the other hand, participants expressed doubt about AI's capabilities to perform usability analysis. One possible reason is that detecting usability problems requires the understanding of the test subject's various behavioral signals and the contexts of the test products and tasks, which was viewed by many participants as too complicated for the AI to fully comprehend.%The reason, we suspect, is that they understood the nuances of and expertise required to analyze usability test videos, given the significant variability of how test subjects expresses concerns over different kinds of system and usability tasks. 

As a result, participants suggested guiding or training the AI to be more similar to how they perform usability analysis.
%, or customized to a given type of usability testing video. 
This suggestion hints that \textit{adaptability} or \textit{customizability} might be a promising area of technical and design innovation for AI-Assisted UX evaluation. Toward this goal, the fields of \textit{active learning}~\cite{settles2009active} and \textit{interactive machine learning}~\cite{fails2003interactive,amershi2014power} provide examples of techniques and systems that allow domain experts without ML expertise to train ML models by supplying examples and tuning the results with interactive interfaces. For example, \textit{explainable active learning}~\cite{teso2019explanatory,Ghai2020XAL} elicits experts' feedback on AI's explanations and incorporates it to train an AI better aligned with how experts make judgments in the task domain. This seems to be a viable approach, as our participants expressed a strong desire to tune the AI, and UX practitioners in general tend to accumulate rich expertise in recognizing the violations of design principles or heuristics that are still too abstract for the AI to pick up from behavioral data alone.
%This could be especially interesting to explore in the context of UX evaluation, as UX professionals are accustomed to the idea of applying design heuristics or rules in performing usability analysis, and our participants expressed strong desire to tune the AI explanations.

Meanwhile, one should be cautious about the common pitfalls of active or interactive machine learning systems~\cite{wu2019local,amershi2014power}, one of which is overfitting, i.e., an AI that fits precisely with expert-provided examples or rules but does not generalize. This is especially problematic as a primary benefit of AI Assistant recognized by our participants was providing a second perspective on their analysis. However, an overfitted AI 
would reinforce their confirmation bias and further trap them in the ``evaluator effect''~\cite{hertzum2001evaluator}. One possible remedy is to explore ensemble approaches with diverse models trained by multiple UX professionals to jointly identify UX problems in usability test videos.

\subsection{Limitations and Future Work}
\label{sec:limitations}

%We used a wizard-of-oz approach to simulating the AI that could identify UX problems and provide explanations for the problems. 
%This approach allowed for studying the research questions without worrying about creating an AI that could detect UX problems for different products.
%Recent research has begun to explore ways to detect UX problems automatically. 
%However, such AIs are often trained and tested on the same set of products~\cite{}. 
\rv{\textbf{WoZ}. While we discussed technical feasibility of implementing different pieces of information in the WoZ AI's explanations in Sec.~\ref{sec:wozAI},
it remains an open question of how to implement such an AI that could generate explanations as good as the WoZ AI. The implication is two-fold. On the one hand, if UX evaluators tend to include AI's explanations into their personal justification, then poor quality or invalidate explanations from such a practical AI could reduce their task performance. On the other hand, we might expect to see lower satisfaction and trust in the AI among participants, who eventually would disengage with it.}

\rv{Our study found that UX evaluators wanted to access more information about AI, such as accuracy and uncertainty. While exposing such information to users might help them understand AI's inferences, it might have downside too. For example, Lim and Dey found that participants could lose their trust in an intelligent agent that shows high uncertainty even if it has a high intelligibility (e.g., being able to explain its reasoning well)~\cite{lim2011investigating}. However, their study was conducted with participants using context-aware applications. It remains unknown how exposing uncertainty and accuracy of AI to UX evaluators would affect their collaboration with the AI.} 
%\cite{fan2020detection,grigera2017automatic,paterno2017customizable,harms2019automated,jeong2020detecting}
%\st{recent research began to explore ways to detect UX problems automatically, it is still challenging to build an AI that is capable of detecting UX problems encountered by different test subjects on a wide range of products. As a result, we used a WoZ approach to simulating the AI that was capable of detecting UX problems for two different products and explaining its inferences to better focus on answering the RQs. While a WoZ AI allows full control over its performance, a real AI might have different nuances in practice, such as making unexpected predictions. Thus, it is worth studying if the findings still hold with a real AI that has a similar performance as the wizard-of-oz AI in the future.}

%The explanations presented a set of features to explain how the AI made its inferences in the current study.
% Second, while participants perceived the set of features in the explanations helpful, they wished to be able to customize the order of the features to better match their expectations and to see other types of information about the AI, such as its accuracy. 
% Consequently, the types of information and their order in the explanations might affect how UX evaluators use and perceive the AI.
% Future work should study the different types of information to explain the AI's inferences and their effect on UX evaluators' performance and preference. 

We simulated the AI with reasonable precision and recall to match the expectation of an imperfect AI in practice. 
Recent research has suggested that the precision and recall of an AI might affect its users' perceptions~\cite{kocielnik2019will}.
Therefore, it \rv{remains an open question of how accessing precision and recall of the AI might affect UX evaluators' collaboration with the AI.}% \st{is worth studying whether the performance of the AI would have an effect on the two factors that we explored in this study. Moreover, further work should explore how the AI's precision and recall might affect UX evaluators' performance.} 
%For example, would a higher false-positive rate or a higher false-negative rate mislead UX evaluators to make more mistakes? 

\rv{\textbf{In-the-Wild Studies}}. As a controlled lab study, we could only include a limited set of UX evaluators and usability test videos. However, the diversity of the products on the market and the diverse backgrounds and experiences of UX evaluators might also play a role in how they use and perceive the AI when they analyze usability test videos. 
More in-the-wild studies with a broader set of products and UX evaluators are needed to further validate the findings.

% Additionally, we set a time-limit for the UX evaluators to analyze each usability test video to ensure that the study could be completed within a time range. 
% We asked participants whether they completed their analysis when the time was up. Among the twenty-four participants, seventeen finished their analysis of the coffee machine session video, and twenty finished their analysis of the website session video.
% In practice, UX evaluators might have a longer time to analyze the videos and thus may adopt a different strategy than our participants used in this study. For example, more time might allow UX evaluators to engage with the AI Assistant more closely, such as scrutinizing and understanding its suggested problems and explanations. This could affect building trust with the AI Assistant in the long run. 

% Moreover, as the UX evaluators analyzed the videos as part of a study, they might have taken different accountability in their analysis than they would when analyzing the videos as part of their daily jobs. 
% Considering these factors, we believe it is worth investigating how UX evaluators would collaborate with the AI and adopt or refuse its suggested problems via an in-the-wild longitudinal study. 

\rv{\textbf{Task Domains Beyond Usability Video Analysis}.} We have studied the \mr{effects} of explanations and synchronization on human-AI collaboration in the context of UX analysis with tasks of analyzing usability test videos. \mr{One should be cautious about generalizing the findings to other task domains. The tasks used in our studies (i.e, identifying usability problems) require expertise in UX research. As a result, our participants were domain experts. Our research and prior work (e.g., ~\cite{fan2020vista,CoUX2021TVCG}) suggested that expert users tended to be confident in their judgements and might have adopted different strategies in human-AI collaboration compared to the crowdworkers or non-experters widely used in human-AI collaboration research. In addition to expertise requirements, the stake of tasks could be another factor affecting the generalization of the findings. While our tasks may be of higher stake than many crowdsourced tasks, much higher stake tasks (e.g., cancer diagnosis) exist.} Thus, although our findings show the positive benefits of explanations and synchronization in fostering an engaging human-AI collaboration, more research is warrant to validate and extend our findings in other task domains \rv{(e.g., tasks requiring expertise vs. tasks without special expertise requirement; high-stake vs low-stake tasks).}

%% file: texfiles/8-conclusion.tex
\section{Conclusion}
%It is important to understand how an AI is presented to knowledge-rich domain workers would affect their performance and preference.
%and thus harvest harvest the potentials of large amounts of test videos collected via remote usability testing.
%In this research, by relating to human-human collaboration, 
We have studied how two factors of human-AI collaboration---explanations and synchronization---would affect the performance and perception of UX evaluators in the context of analyzing usability test videos.
% To study this problem, we iteratively designed a tool---AI Assistant---with four versions that reveal AI-suggested problems and explanations corresponding to the four conditions of the two factors.
%Twenty-four UX evaluators were randomly assigned to use the tool either \textit{with} or \textit{without} explanations. 
%During the study, each UX evaluator analyzed two videos with the \textit{synchronous} and \textit{asynchronous} ways of revealing AI-related information respectively. 
We iteratively designed a tool---AI Assistant---with four versions of UIs corresponding to the two levels of explanations (with/without) and synchronization (sync/async). We conducted a mixed-methods study with 24 UX evaluators identifying UX problems from usability test videos with AI Assistant.

Our quantitative results show that both explanations and synchronization have positive effects on UX evaluators' performance (e.g., the number of identified UX problems) and engagement (e.g., time spent analyzing the video). Specifically, AI with explanations, regardless of being presented synchronously or asynchronously, helped to improve UX evaluators' performance of and engagement in their analysis, as well as perception of the tool (e.g., understanding and satisfaction), compared to the baseline AI (i.e., asynchronous AI without explanations); when without explanations, synchronous AI improved UX evaluators' performance and engagement more than the asynchronous baseline AI.  

Our qualitative results further reveal the ways how explanations and synchronization helped UX evaluators. Specifically, explanations helped them better understand how the AI works, increased their trust in AI, and provided them with additional utility for UX evaluation tasks; and synchronization helped them better perform independent analysis, feel less overwhelmed by the AI's suggestions, and create a stronger sense of ``social presence''---the feeling of working together with a ``colleague.'' 

Based on our findings, we have shown ways to improve the design of explanations and synchronization aspects of human-AI collaboration, such as matching the AI's agency with the type of support it can provide (i.e., AI's capability), balancing between human agency and AI agency, and supporting the adaptability and customizability by allowing UX practitioners to guide the AI, such as by offering their expertise.  
Finally, we have presented the design implications for AI agency and timing of human-AI collaboration in general and for AI-Assisted tools for UX evaluation.

%and highlighted future directions for designing synchronization and explanations of AI, among other factors, to better support human-AI collaboration. 